\documentclass[sigconf]{acmart}
\AtBeginDocument{%
  \providecommand\BibTeX{{%
    \normalfont B\kern-0.5em{\scshape i\kern-0.25em b}\kern-0.8em\TeX}}}

\acmYear{2025}
\setcopyright{none}
\acmConference[]{Conference}{2025}{}
\usepackage{microtype}
\microtypecontext{spacing=nonfrench}

\emergencystretch=\maxdimen
\RequirePackage{caption}

\RequirePackage{tabularx}
\usepackage{multirow}
\usepackage{makecell}

\usepackage{xcolor}
\usepackage{soul}
\newcommand\newtext[1]{#1}
\newcommand\oldtext[1]{}

\graphicspath{{figures/}}

\begin{document}

\title{Talk Me Through It: Developing Effective Systems for Chart Authoring}


\author{Nazar Ponochevnyi}
\email{nazar.ponochevnyi@mail.utoronto.ca}
\orcid{0000-0002-5611-0773}
\affiliation{%
  \department{Faculty of Information} 
  \department{Department of Computer Science} 
  \institution{University of Toronto}
  \city{Toronto}
  \state{Ontario}
  \country{Canada}
}
\author{Young-Ho Kim}
\email{yghokim@younghokim.net}
\orcid{0000-0002-2681-2774}
\affiliation{%
  \institution{NAVER AI Lab}
  \country{Republic of Korea}
}
\author{Joseph Jay Williams
}
\email{williams@cs.toronto.edu}
\orcid{0000-0002-9122-5242}
\affiliation{%
  \department{Department of Computer Science} 
  \institution{University of Toronto}
  \city{Toronto}
  \state{Ontario}
  \country{Canada}
}
\author{Anastasia Kuzminykh}
\email{anastasia.kuzminykh@utoronto.ca}
\orcid{0000-0002-5941-4641}
\affiliation{%
  \department{Faculty of Information} 
  \department{Department of Computer Science} 
  \institution{University of Toronto}
  \city{Toronto}
  \state{Ontario}
  \country{Canada}
}


\begin{abstract}
\oldtext{Recent chart-authoring systems increasingly focus on natural language input but tend to integrate voice capabilities by relying on voice typing, forwarding transcribed voice to the same processor as typed inputs.}
\newtext{Recent chart-authoring systems increasingly focus on natural-language input, enabling users to form a mental image of the chart they wish to create and express this intent using spoken instructions (spoken imagined-chart data).}
\oldtext{However, cross-modality comparisons in other interaction domains suggest that spoken and typed interactions could notably differ.}
\newtext{Yet these systems are predominantly trained on typed instructions written while viewing the target chart (typed existing-chart data). While the cognitive processes for describing an existing chart arguably differ from those for creating a new chart, the structural differences in the corresponding prompts remain underexplored.}
\oldtext{We present empirical findings on the structural differences between spoken and typed instructions for chart creation, showing that spoken prompts have distinctive elements that are not properly captured in a typed-input-oriented analysis pipeline.}
\newtext{We present empirical findings on the structural differences among spoken imagined-chart instructions, typed imagined-chart instructions, and typed existing-chart instructions for chart creation, showing that imagined-chart prompts contain richer command formats, element specifications, and complex linguistic features, especially in spoken instructions.}
\oldtext{We then compare the performance and user satisfaction with chart-authoring systems trained on either spoken or typed data, showing that the voice-trained system matches the performance of the text-trained one on typed input and outperforms it on spoken input, which highlights the necessity of targeted training for voice-enabled systems.}
\newtext{We then compare the performance of systems trained on spoken imagined-chart data versus typed existing-chart data, finding that the first system outperforms the second one on both voice and text input, highlighting the necessity of targeted training on spoken imagined-chart data.}
\oldtext{We conclude with design guidelines for effective voice-based chart-authoring systems and enhancements required for text-based systems to enable spoken input.}
\newtext{We conclude with design guidelines for chart-authoring systems to improve performance in real-world scenarios.}

\end{abstract}




\maketitle

\section{Introduction}

\newtext{The growing popularity of the recent chart-authoring systems, such as Agent in Tableau\footnote{\url{https://www.tableau.com/products/tableau-agent}}, Amazon Q in QuickSight\footnote{\url{https://aws.amazon.com/q}}, and Copilot for Power BI\footnote{\url{https://learn.microsoft.com/en-us/power-bi/create-reports/copilot-introduction}}, 
highlights a growing focus on natural language input for data visualization.}
Furthermore, while the majority of these systems are predominantly focused on text input, there is 
an increasing interest in augmenting text with other input modalities, 
\oldtext{especially voice}\newtext{with a particular interest in voice input} \cite{gpt4o, geminiteam2023gemini}.
Indeed, the integration of voice 
modality provides a lot of promise, for instance, for improving accessibility~\cite{Doush01,Prakash01}, facilitating multi-device setups~\cite{Tucker01,fraser2020remap}, and offering greater freedom of expression~\cite{Badam2017AffordancesOI}. 
Currently, chart-authoring systems tend to integrate voice input capabilities by \newtext{predominantly }relying on
voice typing, i.e., forwarding transcribed spoken queries to the same pipeline as the typed queries~\cite{Setlur01, Gao01, Tang01}. 
The validity of this approach is predicated upon semantic similarities of spoken and typed input in chart-authoring tasks. 
However, cross-modality comparisons of input in other interaction domains, e.g., search~\cite{Guy01,melumad2023vocalizing}, suggest that 
the structure of spoken and typed-in interactions could notably differ, reflecting variations in user expectations based on the interface affordances~\cite{Setlur02}. 
\newtext{Moreover, 
the 
datasets that are employed to train these systems mainly consist of descriptions of the presented charts (i.e., people describe the existing chart) instead of more naturally occurring chart-authoring instructions (i.e., people describe a mental image of a desired chart). 
For example, the 
NLV Corpus dataset \cite{Srinivasan01}, commonly used for training and evaluating natural language interpreters (e.g., Large Language Models) for chart creation \cite{Chen01, Tian2023ChartGPTLL, Sah01, Maddigan2023Chat2VISFD}, was collected by providing ten pairs of tables and charts to 
the participants, asking them to write a description of the presented chart. 
However, in a practical scenario of chart authoring, users most typically form a mental image of their desired visualization and then convey instructions to the chart-authoring system (a process known as prompting or prompt engineering)~\cite{Masry2024ChartInstructIT,gu2023systematicsurveypromptengineering}. While the cognitive processes underlying the descriptions of 
an existing chart arguably differ from those underlying the description of a new chart to be created \cite{Hay_Duffy_McTeague_Pidgeon_Vuletic_Grealy_2017}, the structural differences of the corresponding verbal descriptions remain underexplored.}

The goal of \emph{the first phase} of this work was to explore the semantic alignment\footnote{We define \emph{semantic alignment} as the degree to which the meaning and conceptual structure of user instructions remain consistent\oldtext{ across various input modalities}.} of \oldtext{free-form spoken and typed chart-authoring instructions}\newtext{spoken imagined-chart instructions (voice modality; from mental-image), typed imagined-chart instructions (text modality; from mental-image), and typed existing-chart instructions (text modality; from a presented chart) for chart creation}. 
Through \oldtext{a user study ($N=25$)}\newtext{two user studies}, we 
found that while both \oldtext{text and voice}\newtext{imagined-chart instructions and existing-chart} instructions often cover the basic chart elements and element organization, \oldtext{spoken}\newtext{imagined-chart} instructions have a significantly greater variety of command formats, element characteristics, and complex linguistic features\newtext{, especially in spoken instructions}.

In \emph{the second phase} of this work, we 
developed two comparable systems: 
one trained on 76 transcribed spoken \newtext{imagined-chart} instructions collected in Phase I (System 1) and the 
other on 76 typed existing-chart instructions collected from NLV Corpus \cite{Srinivasan01} (System 2), and compared systems performance \oldtext{and user satisfaction }to determine whether the disparities between spoken \newtext{imagined-chart} and typed \newtext{existing-chart} input for chart creation justify targeted training on the spoken \newtext{imagined-chart instructions} dataset.
Through an evaluation study\oldtext{ ($N=19$)}, we 
found that 
user instructions via spoken \newtext{imagined-chart} input have distinctive elements (e.g., iterative commands, references to the chart elements created earlier, and meta descriptions) that are not properly captured and exploited by System 2. We identified 4 chart aspects where systems often misinterpreted participant commands: chart type, values, design, and annotations. 
\oldtext{While there is no statistically significant difference in the performance of the spoken- and typed-trained systems on typed input, the spoken-trained system outperforms the typed-trained system on spoken input, highlighting the significance of targeted training of chart-authoring systems on the spoken instructions dataset 
to capture the richness of the spoken input.}
\newtext{We found that System 1 outperforms System 2 on both spoken and typed input, highlighting the significance of targeted training of chart-authoring systems on spoken imagined-chart data to capture the richness of more naturally occurring chart-authoring instructions.}

Based on these findings, we discuss the design implications for multimodal systems, including the collection of training datasets in more naturally occurring interactions and choosing or combining input modalities that provide users with the most freedom of expression. 
This work contributes: 
(1) the structure of spoken \newtext{and typed} imagined-chart instructions for chart creation and its similarities and differences with equivalent typed \oldtext{instructions}\newtext{existing-chart instructions from datasets popular for training natural language interpreters};
(2) a set of design implications for \oldtext{developing voice-enabled authoring systems for customizable chart creation}\newtext{multimodal systems to improve performance in real-world scenarios}; and 
(3) a publicly available dataset of spoken \newtext{and typed} \oldtext{descriptions}\newtext{imagined-chart instructions} for chart creation to support the development of \oldtext{spoken-enabled }chart-authoring systems\footnote{
\url{https://github.com/CookieLab-UofT/Voice-Chart-Authoring-Instructions-Dataset}}.  

\section{Related Work}
Incorporating Natural Language Interfaces (NLIs) into visualization authoring systems offers several advantages by removing the requirement for users to have extensive knowledge of a specific interface~\cite{Wang01,Cui01}.
Text input is considered one of the most common modalities for NLIs in chart creation (e.g. ADVISor \cite{Liu01}, Pragmatics Principles \cite{Hoque01}, VisTalk \cite{Wang01}, Snowy \cite{Srinivasan02}, Text2Chart \cite{Rashid01}, etc.). However, 
findings from prior studies indicate that text input can pose accessibility issues \cite{Jung01} and users might struggle to formulate direct commands in text \cite{belkin1980anomalous}.
As a result, there has been a growing interest in developing multimodal interfaces to address the limitations of typed input 
(e.g. Sevi \cite{Tang01}, Eviza \cite{Setlur01}, Analytical Chatbot \cite{Setlur02}, Multi-Modal NLI \cite{Cox01}, VoicePen \cite{Harada01}, DataTone \cite{Gao01}, NL4DV \cite{Arpit01}, etc.).
Training multimodal systems introduces significant challenges. We specifically examine two critical aspects: the role of modality and the nature of data collection tasks.

\subsection{Role of Modality}
Among modalities to augment text input, voice modality has gained a particular interest. For example, \citet{Tang01} developed Sevi, a system that enables users to analyze data through a speech-to-visualization interface. Voice input has the potential to democratize chart creation, 
allowing users to simply speak out loud what they want to see, i.e., provide voice prompts,
potentially improving accessibility \cite{Doush01, 10.1145/3686215.3690154} and enabling multi-device setups \cite{Tucker01}. 
Currently, chart-authoring systems often incorporate voice input by using speech-to-text transcription \cite{Setlur01, Gao01, Tang01, Shen01, Hoque02}. This implies an assumed similarity in how spoken and typed prompts are processed for AI-assisted chart authoring: 
the systems are designed to recognize and execute the same functionalities regardless of whether they are prompted by text or voice. 
However, spoken input is known to have a high degree of freedom of expression \cite{lin2024rambler,mahed2024llms,Aziz01}, constrained only by the user's ability to express a query in natural language \cite{Badam2017AffordancesOI,Gorniak01}, and comparisons of spoken and written input in other interaction domains, e.g., search \cite{Guy01,melumad2023vocalizing}, suggest notable semantic differences between the two. \oldtext{This raises the question of whether the required functionalities expressed through text prompts for chart authoring indeed directly correspond to those expressed through voice prompts.}\newtext{For instance, textual interactions may promote better semantic integration \cite{Chafe01, Bigot01}, while voice queries often resemble natural language more closely \cite{Guy01, melumad2023vocalizing}. Voice queries also tend to include more repetitions \cite{Bigot02, Bigot01} and metacognitive statements that reveal cognitive states \cite{D'Mello01}. However, findings can vary depending on the specific task; for example, some studies indicate that spoken queries are more verbose than written ones \cite{Oviatt01, Bigot03, Bigot01}, while others find no significant differences in length \cite{D'Mello01, Litman01}. These mixed results highlight the importance of investigating cross-modality differences in 
chart authoring.}

\subsection{Role of Data Collection Task}
\newtext{Typically, users form a mental image of their desired creation and then convey instructions to the chart-authoring system (a process known as prompting or prompt engineering)~\cite{gu2023systematicsurveypromptengineering,ROSMI}. 
Forming a mental image of a desired visualization based on the dataset content, without displaying the target chart, more closely resembles the practical scenario of chart authoring \cite{Srinivasan04}.}
\oldtext{To enable the ability of a chart-authoring system to adhere to instructions given in NL prompts, researchers compile visualization specifications into datasets used to train and evaluate machine learning NL interpreters.}\newtext{To enable the ability of a chart-authoring system to adhere to instructions given in prompts, researchers compile visualization specifications into datasets used to train and evaluate natural language (NL) interpreters (e.g., Large Language Models). However, existing datasets used to train these systems mainly consist of descriptions of the presented charts (i.e., people describe the existing chart) instead of more naturally occurring chart-authoring instructions (i.e., people describe a mental image of a desired chart).} 
For example, 
the nvBench dataset \cite{Luo01} (25,750 pairs of descriptions and visualizations) was synthesized to evaluate models in 105 different domains and 7 types of visualizations. 
Previous research examined the structure of NL input to assess how well current tools meet people's expectations for NLIs in data visualization. 
\citet{Setlur01} provided participants with five visualizations and collected various utterances people might use when interacting with a chart. The study identified different tasks people attempted to perform, such as searching, filtering, and changing the chart type, and informed the design of the Eviza system. Similarly, \citet{Wang01} conducted a formative study where participants performed the visualization recreation task and found that people usually start with the chart type and encoding, and then follow up with more specifications. Additionally, participants used data labels, properties, and relations to refer to the objects. Based on these findings, the authors developed the VisTalk tool that allows users to edit visualizations using typed NL input. 
\citet{Srinivasan01} provided ten pairs of tables and charts to 102 participants and asked users to write a text description of the chart to understand how people tend to specify visualization specifications and what types of references they commonly use. The researchers identified four types of phrasing, including commands, queries, questions, and other variations. Additionally, the authors discovered five types of references: attribute, chart type, encoding, aggregation, and design. The NLV Corpus dataset (n=893) 
was publicly released.

\newtext{The validity of training on typed existing-chart instructions (text modality; from a presented chart) is predicated upon semantic similarities of typed existing-chart instructions and spoken imagined-chart instructions (voice modality; from mental-image). However, the structural differences in the corresponding prompts and how training on typed existing-chart instructions affects system performance (i.e., ensuring the system acts upon all user commands correctly) remain underexplored.} 
Building on established practices, we analyze the structural and content aspects of \oldtext{free-form voice instructions for chart creation}\newtext{spoken and typed imagined-chart instructions for chart creation}. 
Then, by investigating their similarities and differences with equivalent \oldtext{text instructions}\newtext{typed existing-chart instructions}, we evaluate the semantic alignment of \oldtext{spoken and written chart-authoring prompts}\newtext{imagined-chart instructions and existing-chart instructions}. 

\section{Phase I - Exploring Semantic Structure of Chart Creation Instructions}
The goal of the first phase was to investigate the semantic similarities and differences between \oldtext{the free-form spoken (voice modality) and typed (text modality) chart-authoring instructions}
\newtext{the spoken imagined-chart instructions (voice modality; from mental-image), typed imagined-chart instructions (text modality; from mental-image), and typed existing-chart instructions (text modality; from a presented chart)}. In this section, we outline the methodology used to gather \oldtext{voice and text instructions}\newtext{the datasets}, compare \oldtext{spoken and typed instructions}\newtext{them}, and present our results.

\subsection{Method}
\subsubsection{Data Collection}
We first composed \oldtext{three}\newtext{four} datasets of chart-authoring instructions: \newtext{spoken imagined-chart instructions} (collected through a user study), \newtext{typed imagined-chart instructions (collected through a user study), } \newtext{typed existing-chart instructions} (from NLV Corpus \cite{Srinivasan01}), and \newtext{synthetic typed existing-chart instructions} (from the nvBench \cite{Luo01}).


\textbf{\oldtext{Voice}\newtext{Spoken Imagined-Chart Instructions} Dataset.} We recruited participants through email lists targeting students, researchers, and individuals who regularly create charts as part of their work. 
Among 25 participants, all were proficient in English; 
13 self-reported as male, 
12 as female. Participants' self-provided prior experience 
creating charts: None=0, Beginner=7, Intermediate=13, Advanced=5; familiarity with conversational assistants: None=11, Infrequent=3, Regular=8, Expert=3.
Participants were provided with text-based, scenario-driven stimuli. \newtext{Text passages of around 40-100 words have been demonstrated to provide sufficient context and attributes for data interpretation tasks \cite{Tandon01}. }To gather authentic examples of such texts, we looked for materials that frequently require visual aids, such as articles, reports, and publications in fields like economics, geography, statistics, demographics, and news. 
From these texts, we only chose those under 100 words; with content on neutral topics (to avoid potential distress for participants); containing ordinary well-known information, which could be supplemented with visualizations — e.g. numerical values or comparisons of concepts. The final corpus 
included 8 texts from 54 to 100 words from Statista \cite{statista} (see Supplemental Materials)
. 

The procedure was approved by the institution's research ethics board and all participants provided informed consent.
The approx. 30-minute moderated study was conducted online via Zoom and audio recorded. Following the instructions, participants were provided with 4 text stimuli, randomly selected from a corpus of 8 texts. 
For each text, participants were asked to imagine a chart to represent the information in it and to describe it verbally as if giving instructions to a virtual assistant on how to create it.
The researchers asked clarifying questions about the participant's instructions throughout the process. 
The study concluded with a semi-structured debriefing interview about the participants' overall experience during the study and their experience working with charts and voice assistants.

Interviews were anonymized, transcribed, and stored on a secure server. 
The final dataset 
included 100 prompts across all texts. 
Given the random assignment of texts to participants, the number of collected instructions per text varied
, with an 
av. of 12.5 $\pm$ 2.29. 
We examined various types of charts included in the voice dataset and found 51 bar, 17 line, 8 scatter, 20 pie, and 4 map charts.
Since bar, line, and scatter charts were also present in both NLV Corpus \cite{Srinivasan01} and nvBench \cite{Luo01} datasets, we narrowed our focus to only voice prompts of these charts (n=76).

\newtext{\textbf{Typed Imagined-Chart Instructions Dataset.} We recruited participants through email lists targeting students, researchers, and individuals who regularly create charts as part of their work. 
Among 23 participants, all were proficient in English; 
6 self-reported as male, 
17 as female. Participants' self-provided prior experience 
creating charts: None=0, Beginner=9, Intermediate=11, Advanced=3; familiarity with conversational assistants: None=2, Infrequent=12, Regular=7, Expert=2.}

\newtext{Participants were provided with text-based, scenario-driven stimuli. We used the same stimuli corpus as for the collection of the spoken imagined-chart instructions dataset. The procedure was approved by the institution's research ethics board and all participants provided informed consent.
The approx. 15-minute moderated study was conducted online via Zoom. Following the instructions, participants were provided with 4 text stimuli, randomly selected from a corpus of 8 texts. 
For each text, participants were asked to imagine a chart to represent the information in it and to type the instructions as if giving instructions to a virtual assistant on how to create it.}

\newtext{The final dataset 
included 92 prompts across all texts. 
Given the random assignment of texts to participants, the number of collected instructions per text varied
, with an 
av. of 11.5 $\pm$ 2.93. 
We examined various types of charts included in the dataset and found 34 bar, 29 line, 2 scatter, 25 pie, and 2 map charts.
Since bar, line, and scatter charts were also present in both NLV Corpus \cite{Srinivasan01} and nvBench \cite{Luo01} datasets, we narrowed our focus to only prompts of these charts (n=65).}

\textbf{\newtext{Typed Existing-Chart Instructions }Datasets.} To compare \oldtext{voice}\newtext{imagined-chart instructions} with the structure of \oldtext{text}\newtext{existing-chart} instructions, we selected only bar, line, and scatter chart types. We randomly selected 
200 typed existing-chart instructions from the NLV Corpus \cite{Srinivasan01}. Additionally, we wanted to compare it with the synthetic typed existing-chart instructions, so we randomly chose 200 synthetic text descriptions from the nvBench \cite{Luo01} dataset. We preserved the same ratio between chart types as in our \newtext{spoken imagined-chart instructions} dataset. 

\subsubsection{Data Analysis}




We analyzed \newtext{spoken imagined-chart} instructions based on two main aspects: how they are phrased (input strategies) and the information they convey on a particular visualization (elements). The average description word count for the \newtext{spoken imagined-chart instructions} dataset -- 175.41 $\pm$ 114.12, \newtext{typed imagined-chart instructions dataset -- 79.95 $\pm$ 47.63, }the NLV Corpus dataset -- 10.06 $\pm$ 4.58, the nvBench dataset -- 25.19 $\pm$ 7.74 (see Supplemental Materials). 
We conducted a Thematic Analysis~\citet{braun2006using} to qualitatively analyze the characteristics of the \newtext{spoken imagined-chart} instructions. Two researchers went through the \newtext{spoken imagined-chart instructions} dataset to develop an initial set of open codes, first focusing on instruction elements; 
the codes were then refined in subsequent passes through the data, 
allowing new codes to emerge and for existing codes to evolve as needed. 
Then, researchers collaboratively identified themes by examining the relationships between codes and refined them as necessary, resulting in 22 instruction elements. Finally, the elements were organized into 5 major types (Fig. \ref{fig:heatmaps}).  
To identify input strategies\oldtext{ in the description}, we applied a top-down approach, starting with 
the coding scheme by \citet{Srinivasan01} and added new codes where needed, identifying 6 input strategies. 
We applied the developed coding scheme to the \newtext{typed imagined-chart instructions and typed existing-chart instructions }datasets and assessed its fitness to identify the structural similarities and differences between the \oldtext{spoken and typed}\newtext{spoken imagined-chart instructions, typed imagined-chart instructions, and typed existing-chart} instructions.

\subsection{Results}

\begin{figure*}
    \centering
    \includegraphics[width=13cm,keepaspectratio]{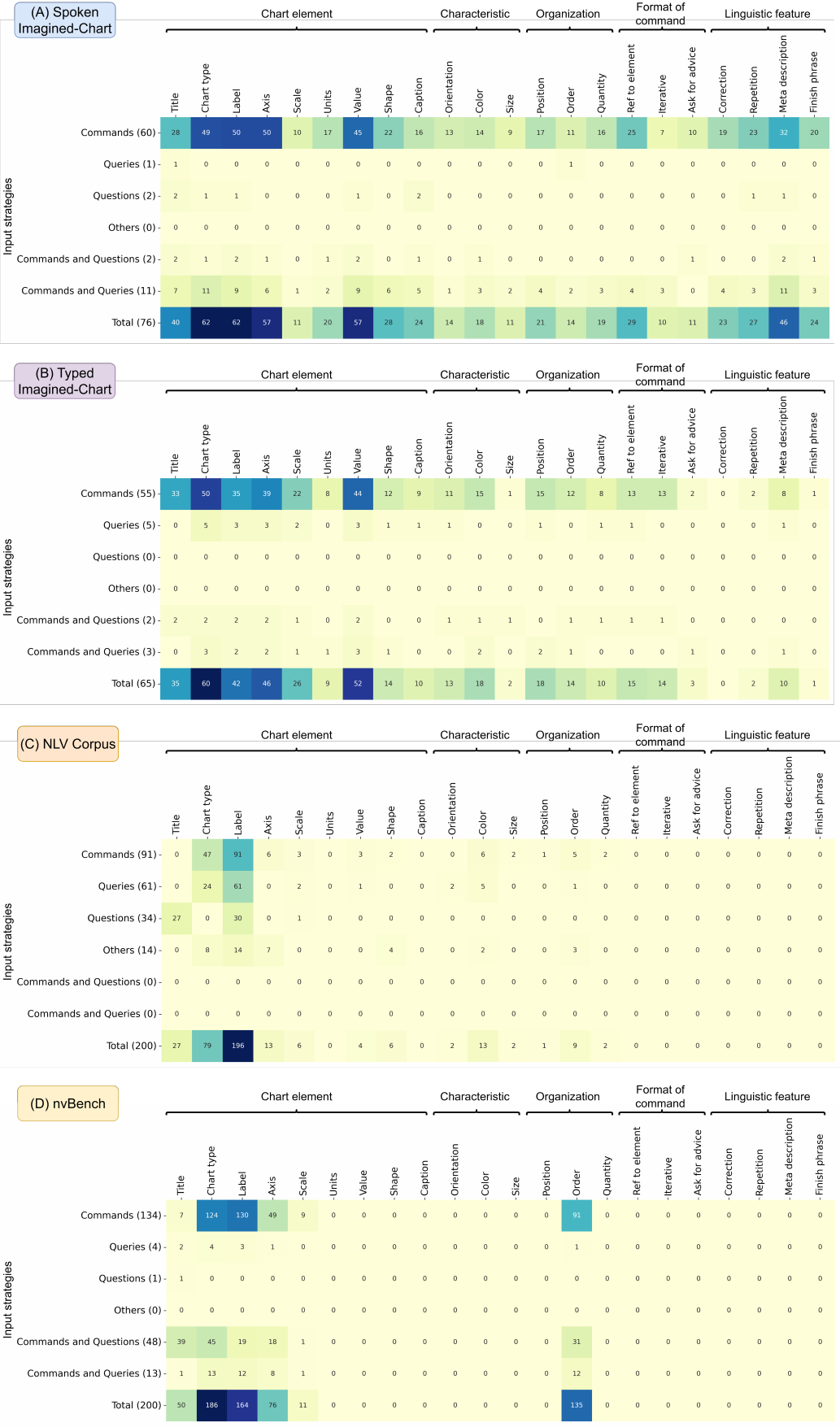}
    \caption{The number of times each element of 5 major types was applied to each input strategy in spoken \newtext{imagined-chart (A) and typed imagined-chart (B) }chart-authoring instructions collected in Phase I, \newtext{typed existing-chart} instructions from the NLV Corpus \cite{Srinivasan01} (C), and synthetic \newtext{typed existing-chart} instructions from the nvBench \cite{Luo01} (D). Color intensity corresponds to data magnitude.}
    \label{fig:heatmaps}
\end{figure*}

Previously, \citet{Srinivasan01} 
identified 4 input strategies common in \newtext{typed existing-chart instructions}: commands (46\% descriptions), queries (30\%), questions (17\%), and others (7\%) (Fig. \ref{fig:heatmaps}.C). Specifically, "commands" are descriptions that sound like requests, "queries" are descriptions that are similar to search queries, "questions" are descriptions that you can answer with a chart, and "others" are descriptions that do not fit either of these categories, e.g. descriptions with special characters that have a programming connotation. In our analysis of spoken \newtext{imagined-chart }instructions, we found that users tend to combine several strategies in one prompt. Thus, we added two additional input strategies: commands and questions (3\%) and commands and queries (14\%) to the existing strategies, such as commands (79\%), queries (1\%), and questions (3\%) (Fig. \ref{fig:heatmaps}.A). 
Synthetic typed existing-chart descriptions also demonstrated some tendencies for combining prompt strategies (Fig. \ref{fig:heatmaps}.D). 

Our qualitative analysis surfaced 5 major types of elements commonly included in \newtext{spoken imagined-chart} instructions: chart element (82\% descriptions), element characteristic (24\%), element organization (28\%), format of command (38\%), and linguistic feature (61\%) (Fig. \ref{fig:heatmaps}.A) 
Specifically, chart elements are the main components of the chart, including the title, type, label, axis, scale, units, values, shape, and caption. 
Element characteristics refer to the attributes of the chart elements, such as orientation, color, and size. 
Element organization refers to how participants arrange the chart elements, including their position, order, and quantity. 
Format of command refers to the way in which participants structure their commands. 
Participants use iterative commands to perform an action for each element, refer to the chart element created earlier, and ask for advice. 
Linguistic features are the ways in which participants use language to describe the chart elements. This includes correcting previously described elements, repeating descriptions of previously mentioned elements, engaging in self-reflection by providing meta descriptions, and finishing the chart description.
We found that participants sometimes use different shapes in the \newtext{spoken imagined-chart instructions} (e.g., lines, asterisks, arrows, icons), and typically provide position and size relative to other elements.
We saw that participants usually did not include a formal finish phrase to signal the completion of their chart description. 
We did not find any significant patterns in the order in which participants provided the various types of elements.
We also found that the general structure for pie and map charts in the \newtext{spoken imagined-chart instructions} dataset was similar to bar, line, and scatter chart types.

After applying the developed coding scheme to the text-based datasets and assessing its fitness, 
\newtext{we found that typed imagined-chart instructions also exhibited "chart element" (98\%), "element characteristic" (46\%), "element organization" (45\%), "format of command" (31\%), and "linguistic feature" (20\%) elements (Fig. \ref{fig:heatmaps}.B). However, "format of command" and "linguistic feature" elements are more often in spoken imagined-chart instructions, so spoken imagined-chart instructions are richer than typed imagined-chart instructions. }
In \newtext{typed existing-chart} instructions, individuals tend to predominantly specify "chart elements" (98\%) (Fig. \ref{fig:heatmaps}.C): label, chart type, implicit title, and axis. However, they rarely mention "element characteristics" (7\%) and "element organization" (5\%).
Synthetic \newtext{typed existing-chart instructions} showcased "chart elements" (93\%) and "element organization" (68\%) only (Fig. \ref{fig:heatmaps}.D). Specifically, we found the following elements
: chart type, label, implicit title, axis, scale, and order. 

To summarize, while both \oldtext{typed and spoken}\newtext{imagined-chart and existing-chart} instructions tend to include basic chart elements and element organization, \oldtext{spoken}\newtext{imagined-chart} prompts have a significantly greater variety of command formats, element characteristics, and complex linguistic features\newtext{, especially in spoken instructions}. 




\section{Phase II - Assessing the Effects of Training Datasets on System Performance}
In this section, we explain how we 
created two comparable systems: the GPT-3.5 Turbo language model fine-tuned on 76 transcribed spoken \newtext{imagined-chart} instructions collected in Phase I (System 1) and the GPT-3.5 Turbo language model fine-tuned on 76 \newtext{typed existing-chart} instructions collected from NLV Corpus \cite{Srinivasan01} (System 2). We then outline the methodology for comparing the performance (i.e., ensuring the system acts upon all user commands correctly) \oldtext{and user satisfaction }of the two systems. Finally, we present our results, 
which show that the differences between \oldtext{spoken}\newtext{imagined-chart} and \oldtext{typed input}\newtext{existing-chart instructions} for creating charts are so significant that they justify targeted training on spoken \newtext{imagined-chart }data.



\subsection{Method}

\subsubsection{System Design}
\textbf{System 1.}
\textit{NL interpreter:} To develop a robust language interpreter that captures the greater structural variability of \newtext{spoken imagined-chart} instructions for chart creation, we decided to use 
a large language model (LLM) trained on a broad dataset from different domains.
Previous research \cite{weng2024insightlensdiscoveringexploringinsights, Voigt2024PlotsMQ, Li2024VisualizationGW, Tian2023ChartGPTLL} demonstrates that LLMs have the potential to recognize and contextually interpret free-form NL input for data visualization. If the user provides context (e.g., table) and instruction to create a chart, 
the model will generate the corresponding chart code. 
Existing literature \cite{liu2022fewshot} indicates that Parameter-efficient fine-tuning (PEFT) offers better accuracy and lower computational costs compared to Few-shot in-context learning (ICL). Thus, we use the former to train the LLM on \oldtext{spoken and typed}\newtext{imagined-chart and existing-chart} datasets.

\textit{Training Dataset:} To adapt LLM to the structure of the spoken \newtext{imagined-chart }chart-authoring instructions, we fine-tune the model \cite{openai-docs} on the dataset of collected transcribed \newtext{spoken imagined-chart instructions} in Phase I.
The fine-tuned LLM can discriminate linguistic features, support various command formats, and determine chart design parameters if the user did not specify them. 
According to the OpenAI documentation \cite{openai-docs}, a dataset of 50 to 100 training examples is recommended to fine-tune the model. Thus, we used 76 spoken \newtext{imagined-chart }instructions from Phase I.
First, two researchers manually created charts for each \newtext{spoken imagined-chart instruction} to get pairs of spoken \newtext{imagined-chart }descriptions and corresponding charts with codes. Next, we compiled a training file with 76 entries: the system message, a user prompt containing text stimulus and transcribed spoken \newtext{imagined-chart }instruction, and an assistant response is a corresponding code to create the desired chart. 

\textit{Code Executor:} 
According to Battle et al. \cite{Battle01}, people increasingly use interactive visualizations for data-driven decision-making.
To create interactive charts, we decided to use Plotly code \cite{plotly2023barcharts}. 
The interactive chart is constructed by executing the LLM-generated code, with the resulting chart showcased on the system's graphical user interface. 
An interactive chart allows users to manipulate the scale, hover over data points to see specific values, and save the graph as a rasterized image.

\textbf{\newtext{System 2}.}
\textit{Training Dataset:} To adapt LLM to the structure of the \newtext{typed existing-chart instructions}, we fine-tune the model \cite{openai-docs} on the NLV Corpus \cite{Srinivasan01}. According to the OpenAI documentation \cite{openai-docs}, it is recommended to use a dataset of 50 to 100 training examples to fine-tune the model. Since we have 76 spoken \newtext{imagined-chart }instructions, we randomly sampled 76 \newtext{typed existing-chart} instructions from a pool of 200 \newtext{typed existing-chart} instructions selected during Phase I. We preserved the same ratio between chart types as in our spoken \newtext{imagined-chart }dataset (51 bar, 17 line, 8 scatter).
NLV Corpus \cite{Srinivasan01} dataset contains \newtext{typed existing-chart} instructions with corresponding chart codes and tables.
First, two researchers manually wrote text stimuli that would describe the values in the tables. We randomly chose up to 12 entries from each table to ensure that the output text stimuli is concise. Next, two researchers manually converted Vega-Lite chart codes \cite{Satyanarayan01} into Plotly codes \cite{plotly2023barcharts} to ensure that \oldtext{typed and spoken}\newtext{imagined-chart and existing-chart} datasets have equivalent pairs of text stimuli and chart codes, and the only significant difference between them is the \oldtext{original modality of the user instructions (voice vs. text)}\newtext{original approach to data collection (people describe a mental image of a desired chart vs. people describe the existing chart)}. Finally, we compiled a training file with 76 entries: the system message, a user prompt containing text stimulus and \newtext{typed existing-chart} instruction, and an assistant response is a corresponding code to create the desired chart.

\textbf{Study environment.}
To evaluate 
two systems for voice and text input, we have developed a web-based testing environment with a minimalistic interface. 
The interface is split into four sections: (1) the information panel on the right displays the current system name and text stimulus; (2) the chart display area on the left shows the history of provided instructions and generated charts for the current text stimulus; (3) the input field on the bottom left is used for recording and typing; (4) the navigation button, located on the bottom right after the user submits the first instruction, allows users to clear the history and proceed to the next text stimulus (see Supplemental Materials). 
We anonymized the system names using codes from A to D. Each system code represents a combination of the system and input modality: (A) \newtext{System 2} with text input, (B) \newtext{System 2} with voice input, (C) \newtext{System 1} with voice input, and (D) \newtext{System 1} with text input. 

\textbf{Implementation.}
We utilized the gpt-3.5-turbo-0125 LLM \cite{GPT35} as a base model for 
both systems to generate Plotly visualization code \cite{plotly2023barcharts}. We fine-tuned the model on \oldtext{typed and spoken}\newtext{imagined-chart and existing-chart} datasets using default hyperparameters (Epochs: 3; Batch size: 1; LR multiplier: 2). For audio transcription, we use the Conformer-1 API \cite{Conformer}. The interactive Gradio interface \cite{gradio} is used to develop the web-based study environment and temporarily deploy it online for the duration of the study. The processing times for text input range from 2 to 10 seconds, and for voice input range from 10 to 60 seconds, depending on the audio recording length. The study environment is implemented in Python.

\subsubsection{Study Design}

\textbf{User Study Participants.}
We recruited participants through email lists targeting students, researchers, and individuals who regularly create charts as part of their work. Among 19 participants, all were proficient in English; 8 self-reported as male, 11 as female. Participants’ self-provided prior experience creating charts: None=1, Beginner=6, Intermediate=9, Advanced=3; familiarity with conversational assistants: None=4, Infrequent=7, Regular=8, Expert=0. 

\textbf{User Study Stimuli Dataset.}
Similar to Phase I, we provided participants with textual scenario-based stimuli. To gather authentic examples of such texts, we searched for four one-variable and four two-variable visualizations from the Statista resource \cite{statista} in economics, geography, statistics, demographics, and news. We only chose those visualizations with captions under 105 words, with content on neutral topics, 
containing ordinary well-known information, which could be supplemented with visualizations. 
The corpus 
included 8 captions from 66 to 105 words (see Supplemental Materials). 

\textbf{User Study Procedure.} The procedure was approved by the institution’s research ethics board and all participants provided informed consent. The approx. 1-hour moderated study was conducted online via Zoom and audio recorded. Participants were asked to share their screen and the screen actions were recorded to document any patterns in the chart creation with the separate consent of the participant. Following the instructions, participants were provided with 8 text stimuli 
and asked to imagine a chart to represent the information in the stimuli and to describe the chart as if giving instructions to a virtual assistant on how to create it. We asked users to use a microphone to provide spoken descriptions for 4 stimuli and a keyboard to provide typed instructions for another 4 stimuli. Participants did not know which system processes the input because we anonymized the system names using codes from A to D. 
We followed the A to D order, repeated twice, but each participant started the study from a random system to make sure that the order of the systems did not affect the outcomes of the study. 
Participants were presented with the chart generated by the system based on their instruction. Upon evaluating the generated chart, participants could make any necessary edits using the same system and answer if they were satisfied with the final result. 
The researchers asked clarifying questions about the participant’s instructions throughout the process. The study concluded with a semi-structured debriefing interview about the participants’ overall experience during the study and their experience working with charts and voice assistants. 

\subsubsection{Data Analysis}
Interviews were anonymized, transcribed, and stored on a secure server. We collected a total of 152 interactions: 76 spoken (38 per system) and 76 typed (38 per system) chart creation interactions across all text stimuli. 
The average number of instructions per interaction was 2.16 $\pm$ 1.21.
In each interaction, we separated the initial (first) instruction provided by the user from the follow-up instructions.
The average initial instruction word count for the spoken instructions – 77.62 $\pm$ 76.76 and for the typed instructions – 57.55 $\pm$ 39.14. 
\oldtext{We applied the coding scheme from Phase I to the initial instructions provided by participants and analyzed the frequency of each type of element (chart elements, element characteristics, element organization, format of command, and linguistic features) in both spoken and typed instructions. }
We \oldtext{then }analyzed the follow-up instructions, distinguishing between instances where participants added new instructions and those where they tried to correct system errors. The coding scheme from Phase I was applied to understand the structure of these correction instructions. 
The analysis of the interactions \oldtext{when the participants were not satisfied with the outcome due to system misunderstandings }resulted in 4 chart aspects where systems often misinterpreted participant commands (Fig. \ref{fig:systems_by_results_checked}). We calculated \oldtext{how often users were satisfied with the final generated chart}\newtext{in how many interactions the system misunderstood the user instructions and failed to include requested elements when generating a chart;} and used the Fisher's Exact Test \cite{Fisher1992} to calculate the statistical significance in the performance of each system on spoken and typed input.

\subsection{Results}
\oldtext{In Phase I, we compared our spoken dataset with an external typed dataset of chart-authoring instructions and found that while both typed and spoken instructions tend to include basic chart elements and element organization, spoken prompts have a significantly greater variety of command formats, element characteristics, and complex linguistic features. 
However, this difference could be attributed to the different study designs used to collect spoken and typed chart-authoring instructions. Thus, we also compared the structure of the spoken and typed instructions collected during the second phase under a controlled evaluation study. For instance, we found that complex linguistic features were present in 41\% of initial and 25\% of correction spoken instructions and in 12\% of initial and 5\% of correction typed instructions, confirming the results of Phase I. }

\begin{figure*}[h]
    \centering
    \includegraphics[width=14cm,keepaspectratio]{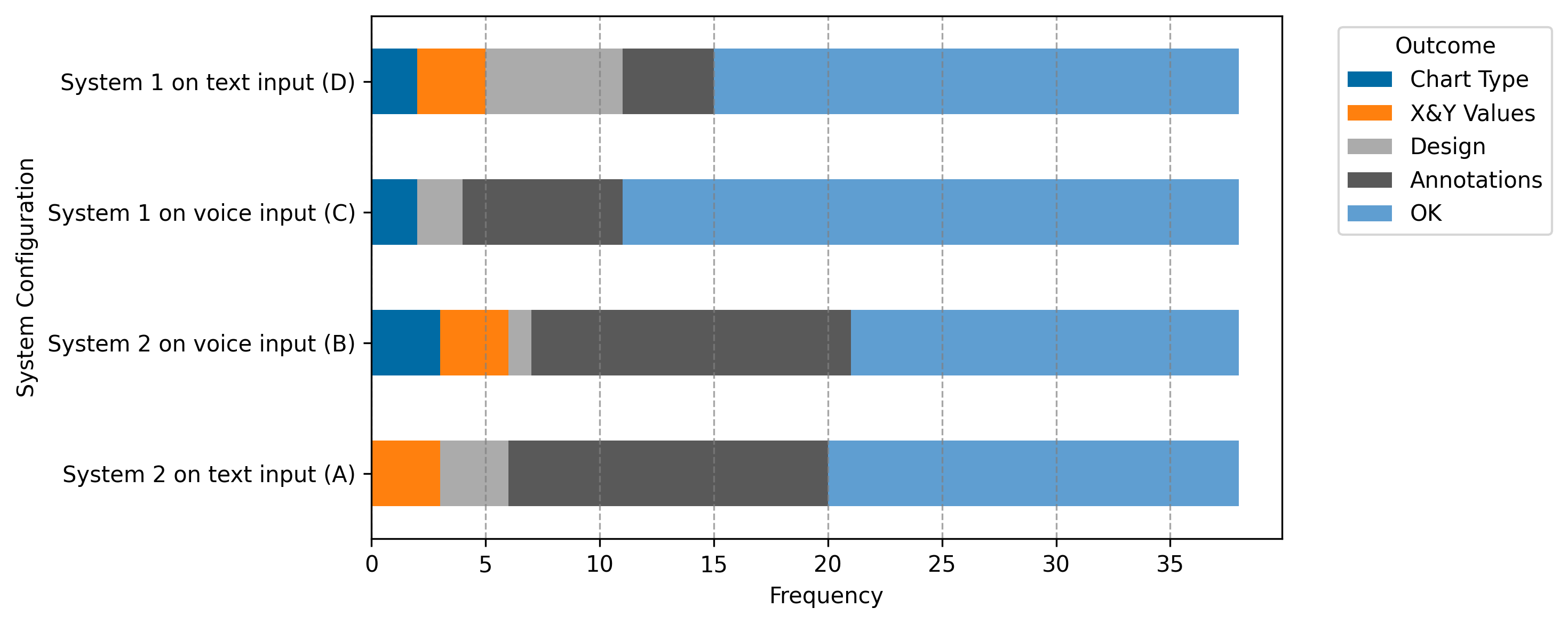}
    \caption{Summary of system performance\oldtext{ and user satisfaction} comparison in Phase II across four configurations: System 2 on text input (A), System 2 on voice input (B), System 1 on voice input (C), and System 1 on text input (D). Each bar represents the frequency of 5 common outcomes: chart type errors, X and Y value errors, design errors, annotation errors, or \oldtext{charts rated as satisfactory by the user}\newtext{correctly generated charts} (OK).}
    \label{fig:systems_by_results_checked}
\end{figure*}

\oldtext{The question remains whether fine-tuning the chart-authoring system on the spoken dataset would improve performance and user satisfaction. }\newtext{We found that training the chart-authoring system on the spoken imagined-chart instructions (i.e., a data collection task that promotes more naturally occurring interactions and a modality that provides more freedom of expression) significantly improves the system performance on voice input: System 1 outperforms System 2 on spoken input }\((p = 0.037 < 0.05)\). System 1 also surpasses System 2 on typed input, but the effects were not statistically significant \((p = 0.357 > 0.05)\). Our analysis of the systems' mistakes surfaced 4 major chart aspects commonly misinterpreted by the systems: chart type, X and Y values, design, and annotations. Chart type errors involved misidentifying the overall structure of the chart, such as confusing bar charts with line charts. Mistakes in X and Y values are incorrect interpretations of data points and axes. Design-related errors are issues with color, order, and size, where systems fail to accurately represent the visual attributes of the chart. Lastly, annotation errors refer to the misinterpretation or omission of captions, text on bars, shapes, and legends. \oldtext{We found that while there is no statistically significant difference in the performance and user satisfaction of the spoken- and typed-trained systems on text input, the spoken-trained system outperforms the typed-trained system on voice input.}
The most common mistake in System 2 on voice input is in annotations (Fig. \ref{fig:systems_by_results_checked}). For example, one participant (P18, System B, Text 5) asked \emph{"Can you add the value labels in the middle of each of the stacks so I can see the exact percentage in each of the value label pairs?"}; however, System 2 misinterpreted the prompt and did not add the value labels. While this is not a critical review of System 2 performance, this example demonstrates how the system trained on typed existing-chart instructions misinterprets the spoken \newtext{imagined-chart }instruction that consists of "chart element", "element organization", "format of command", and "linguistic feature" elements, which, according to Phase I, are commonly found in \newtext{spoken imagined-chart} prompts.

In the semi-structured interview section at the end of the study, participants mentioned the ideal chart-authoring system would first present pre-configured templates and suggest the best chart type based on the data. It would also offer an explanation. Then, the system should support a combination of text, voice, touch, and visual prompts for editing charts. Additionally, the system may proactively ask questions to refine the chart. The system should support visualization code export, remember user preferences, and offer customized suggestions based on prior usage.

\oldtext{To summarize, we found that the spoken-trained system outperforms the typed-trained system on voice input, highlighting the significance of targeted training on spoken input and separate processing of voice instructions to capture the richness of the spoken input}\newtext{To summarize, we found that the system trained on spoken imagined-chart instructions outperforms the system trained on typed existing-chart instructions on both spoken and typed input, highlighting the significance of targeted training on spoken imagined-chart data to capture the richness of more naturally occurring chart-authoring instructions}. 

\section{Discussion}

Recent advancements in multimodal LLMs, such as GPT-4omni from OpenAI and Gemini Live from Google, demonstrate a clear shift toward combining text with other input modalities, particularly voice~\cite{gpt4o, geminiteam2023gemini}. 
Training multimodal systems introduces significant challenges. In this work, we specifically examined two critical aspects: the role of modality and the nature of data collection tasks. 
We advocate for targeted training of multimodal systems on datasets collected in more naturally occurring interactions (e.g., designing a data collection task for training that enables users to form a mental image of the desired output and express that intent in a free-form manner). Moreover, we should choose or combine input modalities that provide users with the most freedom of expression (e.g., allowing users to express their intent in both text and voice modalities during data collection for training). 
However, to effectively choose these training sets and modalities, we must first understand the degrees of distance between different modalities. This necessitates developing measuring approaches and comprehensive descriptions of modality-specific features. Such measures can elucidate the gaps between modalities, guide strategies for filling these gaps through targeted data collection, or determine if training across all available modalities is necessary. Alternatively, they might help identify a single optimal modality that sufficiently encapsulates the benefits of others. 

In the chart-authoring domain, \newtext{most current systems are trained predominantly on typed existing-chart data (i.e., text descriptions of presented charts) because such datasets are easier to collect at scale} and prior work assumed that imagined-chart and existing-chart instructions would not differ meaningfully~\cite{Srinivasan01,Wang01,Luo2025nvBench2A}. 
However, we found that spoken imagined-chart instructions (i.e., descriptions from a mental image of a desired chart) have distinctive elements that are not properly captured and exploited by a system trained on typed existing-chart data. We demonstrated that spoken imagined-chart instructions included all the instruction elements as the typed imagined-chart instructions; and "format of command" and "linguistic feature" elements (e.g., iterative commands, references to the chart elements created earlier, and meta descriptions) are more often in spoken imagined-chart instructions, making them richer than typed imagined-chart instructions. 
Voice input is a more suitable input modality for collecting a training dataset for the chart-authoring system, as it allows for greater freedom of expression~\cite{Badam2017AffordancesOI} and facilitates easy transitions to text input. 
\oldtext{With the assumption that spoken and typed chart-authoring instructions are not significantly different, previous research focused on understanding the structure of the typed input to design NLIs for chart creation. }
\oldtext{They demonstrated that }
\oldtext{chart-authoring systems should }
\oldtext{support natural phrasings (e.g., commands, queries, questions) }
\oldtext{and }
\oldtext{recognize implicit data labels, properties, and references to the objects. }
Spoken instructions 
are generally longer, reflecting a conversational tone and a more verbose communication style, following natural speech patterns more closely than text prompts \cite{Guy01}. 
\oldtext{Exploring efficient ways of handling lengthy input }
\oldtext{is a potential direction for future work. }
Thus, chart-authoring systems should be \oldtext{targeted processing of spoken chart-authoring instructions instead of using typed-input-oriented processing for both typed and spoken prompts}\newtext{trained on spoken imagined-chart data to capture the richness of more naturally occurring chart-authoring instructions}.
\oldtext{developed text-based NL interpreters by fine-tuning LLMs on the dataset of typed and synthetic utterances. }
\oldtext{A similar approach can be adopted for multimodal chart-authoring systems that support both text and voice inputs, i.e., }
\oldtext{by }
\oldtext{fine-tuning the model on datasets for each modality, the system can better recognize and respond to voice and text inputs' unique patterns and interaction structures. }
\oldtext{In the second phase, we created two comparable systems: the GPT-3.5 Turbo language model fine-tuned on 76 transcribed spoken instructions collected in Phase I (spoken-trained system) and the GPT-3.5 Turbo language model fine-tuned on 76 typed instructions collected from NLV Corpus (typed-trained system). The results of the evaluation study showed that the spoken-trained system matches the performance and user satisfaction of the typed-trained one on text input and outperforms it on voice input, highlighting the significance of using a spoken-input-oriented processor for the spoken prompts to capture the richness of the voice input.}
\oldtext{We also found that the spoken-trained system on voice input outperforms the typed-trained system on text input. One possible explanation is that users tend to provide more details through spoken instructions, reducing the commands' ambiguity and leading to more accurate generations. }
\oldtext{Future work could investigate the feasibility of converting transcribed spoken input into typed one, allowing a system trained on typed data to effectively process voice commands. This would involve building an NL translation module using a dataset that includes pairs of spoken instructions and their corresponding typed versions for the same chart.}

Existing-chart datasets are easier to collect at scale because they rely on paired charts and descriptions; however, our findings suggest that relying solely on existing-chart data can miss structurally important aspects of authoring. Future dataset-creation efforts can incorporate imagined-chart collection through practical workflows, such as instrumenting authoring tools to capture natural iterative traces, collecting voice-first intent followed by typed refinements, or eliciting free-form instructions from scenario-based prompts that approximate real authoring goals. Combining such collection with lightweight annotation of action sequences (or deriving them from edit logs) can make imagined-chart data scalable while preserving the richer interaction structures that matter for multimodal authoring performance. 
Our \oldtext{voice dataset}\newtext{imagined-chart voice and text datasets} can 
extend 
current evaluation protocols, like HELM \cite{liang2023holistic} and SuperGLUE \cite{wang2020superglue}. 
This will 
allow researchers to design scenarios that 
resemble real-world applications, ensuring a comprehensive assessment of 
various user input styles.
In particular, evaluation should include multi-turn authoring with corrections, mixed-modality switching, and artifact-referential language (e.g., ``make the third bar darker'', ``move that label''), since these phenomena are frequent in imagined-chart authoring but underrepresented in existing-chart datasets.

\section{Limitations and Future Directions}

Our two-phase study design was focused on the detailed analysis of the structure of 
chart creation instructions and how training data collection task and modality affects system performance. 
Thus, we designed a controlled environment where we can change only a single variable (e.g., \oldtext{input modality}\newtext{data collection task}) and observe how participants would alter the structure of the instructions and how system performance\oldtext{ and user satisfaction} would change. Future research can explore the structure of the \oldtext{multimodal }chart-authoring interactions when adding more modalities \newtext{\cite{Srinivasan03, Bigot01}, highly customized designs \cite{Ren01, Xia01},} and team collaboration \newtext{\cite{Tabalba01}}.

While the controlled environment enabled us to isolate variables and attribute effects more directly, it may not fully capture the complexity of real-world chart authoring (e.g., longer multi-turn sessions, working with messy or unfamiliar datasets, mixed-initiative exploration, switching between modalities, or working within the constraints of specific tools and organizational workflows). Teamwork is common in practice but not strictly required for chart authoring; nevertheless, team dynamics (e.g., negotiation, division of labor, shared context, and iterative handoffs) may affect instruction structure and system error patterns. We hypothesize that collaborative settings may further amplify the richness of imagined-chart instructions relative to existing-chart descriptions, but this requires direct empirical testing and is an important direction for future work.

We focused on text and voice because these are the most common natural-language modalities supported by current chart-authoring tools. At the same time, there is growing interest in additional modalities for authoring and refinement (e.g., touch, pen, gaze, and multimodal combinations), and future work should evaluate how these modalities interact with both instruction-elicitation tasks and input modality, and whether they shift instruction structure, ambiguity, and correction behavior.

\oldtext{Our approach was centered on a detailed manual analysis; thus, we chose a moderated study format. This method allowed us to gain a comprehensive understanding of the structure of the provided instructions and categorize the description elements into five categories.}
\newtext{Our approach was centered on a detailed manual analysis; thus, we chose a moderated study format. This method allowed us to interpret free-form instructions with higher fidelity (e.g., via think-aloud and clarifying questions) and to connect specific utterance structures to intended chart semantics, while still enabling a systematic categorization of instruction elements.}
\newtext{However, moderation can shape how participants phrase instructions; future work should validate our findings in less-instrumented settings, including unmoderated or in-the-wild deployments, and compare instruction structure and error patterns across study formats.}
\newtext{At the same time, we note an increasing trend toward integrating agentic assistants into everyday software workflows, where users may naturally externalize intent in a conversational manner; such contexts may partially resemble the guided interaction style of moderated sessions.}
\oldtext{However, the dataset size was limited compared to other online studies, which collected several hundred utterances. Thus, future dataset expansion can contribute to more effective model training and \oldtext{representation of participants from diverse backgrounds}\newtext{multilingual support}.}
\newtext{Finally, the dataset size was limited compared to other online studies that collect several hundred utterances. }
We believe that expanding the dataset will enhance the observed performance improvements and statistical significance for both voice and text modalities. 
Future dataset expansion can also stratify collection by chart complexity and interaction stage (initial authoring vs. iterative refinement) to better model long, real-world authoring workflows.

\section{Conclusion}
This research highlights the importance of designing targeted data collection tasks for training chart-authoring systems. Our analysis revealed significant structural differences between imagined-chart instructions, especially spoken, and traditional typed existing-chart descriptions. Consequently, systems trained on spoken imagined-chart data outperform those trained on typed existing-chart data on both voice and text input. Future multimodal systems should employ training datasets derived from more realistic data collection tasks and richer modalities to enhance system performance.

\section*{Safe and Responsible Innovation Statement}
Ethical considerations, including user privacy, involved anonymizing participant data and securely storing recordings. We mitigated bias by recruiting diverse participants with various levels of chart-authoring expertise. We acknowledge risks like data misinterpretation and clearly communicate system limitations to prevent misuse. Our open-sourced dataset promotes transparency and accountability in future deployments.

\bibliographystyle{ACM-Reference-Format}
\bibliography{sample-base}

@article{melumad2023vocalizing,
  title={Vocalizing search: How voice technologies alter consumer search processes and satisfaction},
  author={Melumad, Shiri},
  journal={Journal of Consumer Research},
  pages={ucad009},
  year={2023},
  publisher={Oxford University Press}
}

@article{mahed2024llms,
  title={Are LLMs Robust for Spoken Dialogues?},
  author={Mahed Mousavi, Seyed and Roccabruna, Gabriel and Alghisi, Simone and Rizzoli, Massimo and Ravanelli, Mirco and Riccardi, Giuseppe},
  journal={arXiv e-prints},
  pages={arXiv--2401},
  year={2024}
}

@article{lin2024rambler,
  title={Rambler: Supporting Writing With Speech via LLM-Assisted Gist Manipulation},
  author={Lin, Susan and Warner, Jeremy and Zamfirescu-Pereira, JD and Lee, Matthew G and Jain, Sauhard and Huang, Michael Xuelin and Lertvittayakumjorn, Piyawat and Cai, Shanqing and Zhai, Shumin and Hartmann, Bj{\"o}rn and others},
  journal={arXiv preprint arXiv:2401.10838},
  year={2024}
}

@inproceedings{fraser2020remap,
  title={ReMap: Lowering the barrier to help-seeking with multimodal search},
  author={Fraser, C Ailie and Markel, Julia M and Basa, N James and Dontcheva, Mira and Klemmer, Scott},
  booktitle={Proceedings of the 33rd Annual ACM Symposium on User Interface Software and Technology},
  pages={979--986},
  year={2020}
}

@article{braun2006using,
  title={Using thematic analysis in psychology},
  author={Braun, Virginia and Clarke, Victoria},
  journal={Qualitative research in psychology},
  volume={3},
  number={2},
  pages={77--101},
  year={2006},
  publisher={Taylor \& Francis}
}

@String{Computing = "Computing" }

@String{Computer = "{IEEE} Computer" }

@String{Springer = "Springer-Verlag" }

@ARTICLE{Battle01,  author={Battle, Leilani and Scheidegger, Carlos},  journal={IEEE Transactions on Visualization and Computer Graphics},   title={A Structured Review of Data Management Technology for Interactive Visualization and Analysis},   year={2021},  volume={27},  number={2},  pages={1128-1138},  doi={10.1109/TVCG.2020.3028891}}

@ARTICLE{Jung01,  author={Jung, Crescentia and Mehta, Shubham and Kulkarni, Atharva and Zhao, Yuhang and Kim, Yea-Seul},  journal={IEEE Transactions on Visualization and Computer Graphics},   title={Communicating Visualizations without Visuals: Investigation of Visualization Alternative Text for People with Visual Impairments},   year={2022},  volume={28},  number={1},  pages={1095-1105},  doi={10.1109/TVCG.2021.3114846}}

@article{Shen01,
author = {Shen, Leixian and Shen, Enya and Luo, Yuyu and Yang, Xiaocong and Hu, Xuming and Zhang, Xiongshuai and Tai, Zhiwei and Wang, Jianmin},
title = {Towards Natural Language Interfaces for Data Visualization: A Survey},
year = {2023},
issue_date = {June 2023},
publisher = {IEEE Educational Activities Department},
address = {USA},
volume = {29},
number = {6},
issn = {1077-2626},
url = {https://doi.org/10.1109/TVCG.2022.3148007},
doi = {10.1109/TVCG.2022.3148007},
journal = {IEEE Transactions on Visualization and Computer Graphics},
month = jun,
pages = {3121–3144},
numpages = {24}
}

@inproceedings{Rashid01,
  title={Text2Chart: A Multi-Staged Chart Generator from Natural Language Text},
  author={Md. Mahinur Rashid and Hasin Kawsar Jahan and Annysha Huzzat and Riyasaat Ahmed Rahul and Tamim Bin Zakir and Farhana Firoz Meem and Md. Saddam Hossain Mukta and Swakkhar Shatabda},
  booktitle={PAKDD},
  year={2022}
}

@inproceedings{Tang01,
author = {Tang, Jiawei and Luo, Yuyu and Ouzzani, Mourad and Li, Guoliang and Chen, Hongyang},
title = {Sevi: Speech-to-Visualization through Neural Machine Translation},
year = {2022},
isbn = {9781450392495},
publisher = {Association for Computing Machinery},
address = {New York, NY, USA},
url = {https://doi.org/10.1145/3514221.3520150},
doi = {10.1145/3514221.3520150},
booktitle = {Proceedings of the 2022 International Conference on Management of Data},
pages = {2353–2356},
numpages = {4},
location = {Philadelphia, PA, USA},
series = {SIGMOD '22}
}

@online{gradio, author = {Gradio}, title = {Build and Share Delightful Machine Learning Apps}, publisher = {Gradio}, year = {2021}, url = {https://gradio.app} }

@article{Cui01,
  title={Text-to-Viz: Automatic Generation of Infographics from Proportion-Related Natural Language Statements},
  author={Weiwei Cui and Xiaoyu Zhang and Yun Wang and He Huang and B. Chen and Lei Fang and Haidong Zhang and Jian-Guang Lou and Dongmei Zhang},
  journal={IEEE Transactions on Visualization and Computer Graphics},
  year={2020},
  volume={26},
  pages={906-916}
}

@article{Arpit01,
  title={NL4DV: A Toolkit for Generating Analytic Specifications for Data Visualization from Natural Language Queries},
  author={Arpit Narechania and Arjun Srinivasan and John T. Stasko},
  journal={IEEE Transactions on Visualization and Computer Graphics},
  year={2021},
  volume={27},
  pages={369-379}
}

@article{Satyanarayan01,
author = {Satyanarayan, Arvind and Moritz, Dominik and Wongsuphasawat, Kanit and Heer, Jeffrey},
title = {Vega-Lite: A Grammar of Interactive Graphics},
year = {2017},
issue_date = {January 2017},
publisher = {IEEE Educational Activities Department},
address = {USA},
volume = {23},
number = {1},
issn = {1077-2626},
url = {https://doi.org/10.1109/TVCG.2016.2599030},
doi = {10.1109/TVCG.2016.2599030},
journal = {IEEE Transactions on Visualization and Computer Graphics},
month = {jan},
pages = {341–350},
numpages = {10}
}

@article{Liu01,
  title={ADVISor: Automatic Visualization Answer for Natural-Language Question on Tabular Data},
  author={Can Liu and Yun Han and Ruike Jiang and Xiaoru Yuan},
  journal={2021 IEEE 14th Pacific Visualization Symposium (PacificVis)},
  year={2021},
  pages={11-20}
}

@article{Srinivasan01,
  title={Collecting and Characterizing Natural Language Utterances for Specifying Data Visualizations},
  author={Arjun Srinivasan and Nikhila Nyapathy and Bongshin Lee and Steven Mark Drucker and John T. Stasko},
  journal={Proceedings of the 2021 CHI Conference on Human Factors in Computing Systems},
  year={2021}
}

@article{Cox01,
author = {Cox, Kenneth and Grinter, Rebecca and Hibino, Stacie and Jagadeesan, Lalita and Mantilla, David},
year = {2001},
month = {07},
pages = {297-314},
title = {A Multi-Modal Natural Language Interface to an Information Visualization Environment},
volume = {4},
journal = {International Journal of Speech Technology},
doi = {10.1023/A:1011368926479}
}

@article{Luo01,
  title={nvBench: A Large-Scale Synthesized Dataset for Cross-Domain Natural Language to Visualization Task},
  author={Yuyu Luo and Jiawei Tang and Guoliang Li},
  journal={ArXiv},
  year={2021},
  volume={abs/2112.12926}
}

@article{Doush01,
author = {Abu Doush, Iyad and Pontelli, Enrico and Son, Tran Cao and Simon, Dominic and Ma, Ou},
title = {Multimodal Presentation of Two-Dimensional Charts: An Investigation Using Open Office XML and Microsoft Excel},
year = {2010},
issue_date = {November 2010},
publisher = {Association for Computing Machinery},
address = {New York, NY, USA},
volume = {3},
number = {2},
issn = {1936-7228},
url = {https://doi.org/10.1145/1857920.1857925},
doi = {10.1145/1857920.1857925},
journal = {ACM Trans. Access. Comput.},
month = {nov},
articleno = {8},
numpages = {50}
}

@inproceedings{Srinivasan02,
author = {Srinivasan, Arjun and Setlur, Vidya},
title = {Snowy: Recommending Utterances for Conversational Visual Analysis},
year = {2021},
isbn = {9781450386357},
publisher = {Association for Computing Machinery},
address = {New York, NY, USA},
url = {https://doi.org/10.1145/3472749.3474792},
doi = {10.1145/3472749.3474792},
booktitle = {The 34th Annual ACM Symposium on User Interface Software and Technology},
pages = {864–880},
numpages = {17},
location = {Virtual Event, USA},
series = {UIST '21}
}

@misc{openai-docs,
  author = "OpenAI",
  title = "{Fine-Tuning - OpenAI API}",
  year = "2024",
  howpublished = "\url{https://platform.openai.com/docs/guides/fine-tuning}",
    note = {Accessed: 2024-09-04}
}

@misc{statista,
  author = "Statista",
  title = "{Statista - The Statistics Portal for Market Data, Market Research and Market Studies}",
  year = "2007",
  howpublished = "\url{https://www.statista.com}",
  note = "[Accessed: April 17, 2023]"
}

@misc{plotly2023barcharts,
    author = {Plotly},
    title = {Plotly Python Graphing Library},
    year = {2024},
    url = {https://plotly.com/python/},
    note = {Accessed: 2024-09-04}
}

@misc{GPT35,
    author = {OpenAI},
    title = {GPT-3.5 Turbo - OpenAI API},
    year = {2024},
    url = {https://platform.openai.com/docs/models/gpt-3-5-turbo},
    note = {Accessed: 2024-09-04}
}

@article{Conformer,
   title={Conformer: Convolution-augmented Transformer for Speech Recognition},
   url={http://dx.doi.org/10.21437/interspeech.2020-3015},
   DOI={10.21437/interspeech.2020-3015},
   journal={Interspeech 2020},
   publisher={ISCA},
   author={Gulati, Anmol and Qin, James and Chiu, Chung-Cheng and Parmar, Niki and Zhang, Yu and Yu, Jiahui and Han, Wei and Wang, Shibo and Zhang, Zhengdong and Wu, Yonghui and Pang, Ruoming},
   year={2020},
   month={Oct} }

@ARTICLE{Hoque01,
  author={Hoque, Enamul and Setlur, Vidya and Tory, Melanie and Dykeman, Isaac},
  journal={IEEE Transactions on Visualization and Computer Graphics}, 
  title={Applying Pragmatics Principles for Interaction with Visual Analytics}, 
  year={2018},
  volume={24},
  number={1},
  pages={309-318},
  doi={10.1109/TVCG.2017.2744684}}

@inproceedings{Setlur01,
author = {Setlur, Vidya and Battersby, Sarah E. and Tory, Melanie and Gossweiler, Rich and Chang, Angel X.},
title = {Eviza: A Natural Language Interface for Visual Analysis},
year = {2016},
isbn = {9781450341899},
publisher = {Association for Computing Machinery},
address = {New York, NY, USA},
url = {https://doi.org/10.1145/2984511.2984588},
doi = {10.1145/2984511.2984588},
booktitle = {Proceedings of the 29th Annual Symposium on User Interface Software and Technology},
pages = {365–377},
numpages = {13},
location = {Tokyo, Japan},
series = {UIST '16}
}

@ARTICLE{Wang01,
  author={Wang, Yun and Hou, Zhitao and Shen, Leixian and Wu, Tongshuang and Wang, Jiaqi and Huang, He and Zhang, Haidong and Zhang, Dongmei},
  journal={IEEE Transactions on Visualization and Computer Graphics}, 
  title={Towards Natural Language-Based Visualization Authoring}, 
  year={2023},
  volume={29},
  number={1},
  pages={1222-1232},
  doi={10.1109/TVCG.2022.3209357}}

@article{belkin1980anomalous,
  title={Anomalous states of knowledge as a basis for information retrieval},
  author={Belkin, Nicholas J},
  journal={Canadian journal of information science},
  volume={5},
  number={1},
  pages={133--143},
  year={1980}
}

@inproceedings{Badam2017AffordancesOI,
  title={Affordances of Input Modalities for Visual Data Exploration in Immersive Environments},
  author={Sriram Karthik Badam and Arjun Srinivasan and Niklas Elmqvist},
  year={2017},
  url={https://api.semanticscholar.org/CorpusID:20980425}
}

@inproceedings{Gao01,
author = {Gao, Tong and Dontcheva, Mira and Adar, Eytan and Liu, Zhicheng and Karahalios, Karrie G.},
title = {DataTone: Managing Ambiguity in Natural Language Interfaces for Data Visualization},
year = {2015},
isbn = {9781450337793},
publisher = {Association for Computing Machinery},
address = {New York, NY, USA},
url = {https://doi.org/10.1145/2807442.2807478},
doi = {10.1145/2807442.2807478},
booktitle = {Proceedings of the 28th Annual ACM Symposium on User Interface Software \& Technology},
pages = {489–500},
numpages = {12},
location = {Charlotte, NC, USA},
series = {UIST '15}
}

@inproceedings{Setlur02,
author = {Setlur, Vidya and Tory, Melanie},
title = {How Do You Converse with an Analytical Chatbot? Revisiting Gricean Maxims for Designing Analytical Conversational Behavior},
year = {2022},
isbn = {9781450391573},
publisher = {Association for Computing Machinery},
address = {New York, NY, USA},
url = {https://doi.org/10.1145/3491102.3501972},
doi = {10.1145/3491102.3501972},
booktitle = {Proceedings of the 2022 CHI Conference on Human Factors in Computing Systems},
articleno = {29},
numpages = {17},
location = {New Orleans, LA, USA},
series = {CHI '22}
}

@article{Ren01,
author = {Ren, Donghao and Lee, Bongshin and Brehmer, Matthew},
title = {Charticulator: Interactive Construction of Bespoke Chart Layouts},
year = {2019},
issue_date = {Jan. 2019},
publisher = {IEEE Educational Activities Department},
address = {USA},
volume = {25},
number = {1},
issn = {1077-2626},
url = {https://doi.org/10.1109/TVCG.2018.2865158},
doi = {10.1109/TVCG.2018.2865158},
journal = {IEEE Transactions on Visualization and Computer Graphics},
month = {jan},
pages = {789–799},
numpages = {11}
}

@inproceedings{Xia01,
author = {Xia, Haijun and Henry Riche, Nathalie and Chevalier, Fanny and De Araujo, Bruno and Wigdor, Daniel},
title = {DataInk: Direct and Creative Data-Oriented Drawing},
year = {2018},
isbn = {9781450356206},
publisher = {Association for Computing Machinery},
address = {New York, NY, USA},
url = {https://doi.org/10.1145/3173574.3173797},
doi = {10.1145/3173574.3173797},
booktitle = {Proceedings of the 2018 CHI Conference on Human Factors in Computing Systems},
pages = {1–13},
numpages = {13},
location = {Montreal QC, Canada},
series = {CHI '18}
}

@misc{gpt4o,
    author = {OpenAI},
    title = {Hello GPT-4o},
    year = {2024},
    note = {Accessed: 2024-09-04},
    url = {https://openai.com/index/hello-gpt-4o},
}

@inproceedings{Srinivasan03,
author = {Srinivasan, Arjun and Lee, Bongshin and Henry Riche, Nathalie and Drucker, Steven M. and Hinckley, Ken},
title = {InChorus: Designing Consistent Multimodal Interactions for Data Visualization on Tablet Devices},
year = {2020},
isbn = {9781450367080},
publisher = {Association for Computing Machinery},
address = {New York, NY, USA},
url = {https://doi.org/10.1145/3313831.3376782},
doi = {10.1145/3313831.3376782},
booktitle = {Proceedings of the 2020 CHI Conference on Human Factors in Computing Systems},
pages = {1–13},
numpages = {13},
location = {Honolulu, HI, USA},
series = {CHI '20}
}

@article{Tucker01,
author = {Philip Tucker and Dylan M. Jones},
title = {Voice as interface: An overview},
journal = {International Journal of Human–Computer Interaction},
volume = {3},
number = {2},
pages = {145-170},
year  = {1991},
publisher = {Taylor & Francis},
doi = {10.1080/10447319109526002},
URL = { 
        https://doi.org/10.1080/10447319109526002
},
eprint = { 
        https://doi.org/10.1080/10447319109526002
}
}

@misc{geminiteam2023gemini,
      title={Gemini: A Family of Highly Capable Multimodal Models}, 
      author={Gemini Team et al.},
      year={2023},
      eprint={2312.11805},
      archivePrefix={arXiv},
      primaryClass={cs.CL}
}

@article{Guy01,
author = {Guy, Ido},
title = {The Characteristics of Voice Search: Comparing Spoken with Typed-in Mobile Web Search Queries},
year = {2018},
issue_date = {July 2018},
publisher = {Association for Computing Machinery},
address = {New York, NY, USA},
volume = {36},
number = {3},
issn = {1046-8188},
url = {https://doi.org/10.1145/3182163},
doi = {10.1145/3182163},
journal = {ACM Trans. Inf. Syst.},
month = {mar},
articleno = {30},
numpages = {28}
}

@misc{liang2023holistic,
      title={Holistic Evaluation of Language Models}, 
      author={Percy Liang et al.},
      year={2023},
      eprint={2211.09110},
      archivePrefix={arXiv},
      primaryClass={cs.CL}
}

@misc{wang2020superglue,
      title={SuperGLUE: A Stickier Benchmark for General-Purpose Language Understanding Systems}, 
      author={Alex Wang and Yada Pruksachatkun and Nikita Nangia and Amanpreet Singh and Julian Michael and Felix Hill and Omer Levy and Samuel R. Bowman},
      year={2020},
      eprint={1905.00537},
      archivePrefix={arXiv},
      primaryClass={cs.CL}
}

@Inbook{Fisher1992,
author="Fisher, R. A.",
editor="Kotz, Samuel
and Johnson, Norman L.",
title="Statistical Methods for Research Workers",
bookTitle="Breakthroughs in Statistics: Methodology and Distribution",
year="1992",
publisher="Springer New York",
address="New York, NY",
pages="66--70",
isbn="978-1-4612-4380-9",
doi="10.1007/978-1-4612-4380-9_6",
url="https://doi.org/10.1007/978-1-4612-4380-9_6"
}

@misc{weng2024insightlensdiscoveringexploringinsights,
      title={InsightLens: Discovering and Exploring Insights from Conversational Contexts in Large-Language-Model-Powered Data Analysis}, 
      author={Luoxuan Weng and Xingbo Wang and Junyu Lu and Yingchaojie Feng and Yihan Liu and Wei Chen},
      year={2024},
      eprint={2404.01644},
      archivePrefix={arXiv},
      primaryClass={cs.HC},
      url={https://arxiv.org/abs/2404.01644}, 
}

@inproceedings{Voigt2024PlotsMQ,
  title={Plots Made Quickly: An Efficient Approach for Generating Visualizations from Natural Language Queries},
  author={Henrik Voigt and Kai Lawonn and Sina Zarrie{\ss}},
  booktitle={International Conference on Language Resources and Evaluation},
  year={2024},
  url={https://api.semanticscholar.org/CorpusID:269804810}
}

@article{Li2024VisualizationGW,
  title={Visualization Generation with Large Language Models: An Evaluation},
  author={Guozheng Li and Xinyu Wang and Gerile Aodeng and Shunyuan Zheng and Yu Zhang and Chuangxin Ou and Song Wang and Chi Harold Liu},
  journal={ArXiv},
  year={2024},
  volume={abs/2401.11255},
  url={https://api.semanticscholar.org/CorpusID:267069400}
}

@article{Tian2023ChartGPTLL,
  title={ChartGPT: Leveraging LLMs to Generate Charts from Abstract Natural Language},
  author={Yuan Tian and Weiwei Cui and Dazhen Deng and Xinjing Yi and Yurun Yang and Haidong Zhang and Yingcai Wu},
  journal={IEEE transactions on visualization and computer graphics},
  year={2023},
  volume={PP},
  url={https://api.semanticscholar.org/CorpusID:265019126}
}

@inproceedings{
liu2022fewshot,
title={Few-Shot Parameter-Efficient Fine-Tuning is Better and Cheaper than In-Context Learning},
author={Haokun Liu and Derek Tam and Muqeeth Mohammed and Jay Mohta and Tenghao Huang and Mohit Bansal and Colin Raffel},
booktitle={Advances in Neural Information Processing Systems},
editor={Alice H. Oh and Alekh Agarwal and Danielle Belgrave and Kyunghyun Cho},
year={2022},
url={https://openreview.net/forum?id=rBCvMG-JsPd}
}

@inproceedings{Tandon01,
author = {Tandon, Sara and Abdul-Rahman, Alfie and Borgo, Rita},
title = {Visual Task Performance and Spatial Abilities: An Investigation of Artists and Mathematicians},
year = {2023},
isbn = {9781450394215},
publisher = {Association for Computing Machinery},
address = {New York, NY, USA},
url = {https://doi.org/10.1145/3544548.3580765},
doi = {10.1145/3544548.3580765},
booktitle = {Proceedings of the 2023 CHI Conference on Human Factors in Computing Systems},
articleno = {806},
numpages = {16},
location = {Hamburg, Germany},
series = {CHI '23}
}

@inproceedings{Tabalba01,
author = {Tabalba, Roderick S and Kirshenbaum, Nurit and Leigh, Jason and Bhattacharya, Abari and Grosso, Veronica and Di Eugenio, Barbara and Johnson, Andrew E and Zellner, Moira},
title = {An Investigation into an Always Listening Interface to Support Data Exploration},
year = {2023},
isbn = {9798400701061},
publisher = {Association for Computing Machinery},
address = {New York, NY, USA},
url = {https://doi.org/10.1145/3581641.3584079},
doi = {10.1145/3581641.3584079},
booktitle = {Proceedings of the 28th International Conference on Intelligent User Interfaces},
pages = {128–141},
numpages = {14},
location = {Sydney, NSW, Australia},
series = {IUI '23}
}

@incollection{Chafe01,
  author    = {Chafe, Wallace L.},
  title     = {Integration and involvement in speaking, writing, and oral literature},
  booktitle = {Spoken and Written Language},
  editor    = {Tannen, Deborah},
  volume    = {IX},
  pages     = {35--53},
  year      = {1982},
  publisher = {Ablex Publishing Corporation},
  address   = {Norwood, New Jersey}
}

@article{Bigot01,
  author    = {{Le Bigot}, Ludovic and Terrier, Patrice and Jamet, Eric and Botherel, Val{\'e}rie and Rouet, Jean-Fran{\c c}ois},
  title     = {Does textual feedback hinder spoken interaction in natural language?},
  journal   = {Ergonomics},
  volume    = {53},
  number    = {1},
  pages     = {43--55},
  year      = {2010},
  publisher = {Taylor \& Francis},
  doi       = {10.1080/00140130903306666},
  note      = {PMID: 20069480},
  url       = {https://doi.org/10.1080/00140130903306666},
  eprint    = {https://doi.org/10.1080/00140130903306666}
}

@article{Bigot02,
  TITLE = {{Effect of modality on collaboration with a dialogue system}},
  AUTHOR = {Le Bigot, Ludovic and Terrier, Patrice and Amiel, Virginie and Poulain, G{\'e}rard and Jamet, Eric and Rouet, Jean-Fran{\c c}ois},
  URL = {https://hal.science/hal-01806640},
  JOURNAL = {{International Journal of Human-Computer Studies}},
  PUBLISHER = {{Elsevier}},
  VOLUME = {65},
  NUMBER = {12},
  PAGES = {983 - 991},
  YEAR = {2007},
  MONTH = Dec,
  DOI = {10.1016/j.ijhcs.2007.07.002},
  HAL_ID = {hal-01806640},
  HAL_VERSION = {v1},
}

@article{Bigot03,
author = {Le Bigot, Ludovic and Jamet, Éric and Rouet, Jean-François and Amiel, Virginie},
year = {2013},
month = {03},
pages = {467-500},
title = {Mode and modal transfer effects on performance and discourse organization with an information retrieval dialogue system in natural language},
journal = {Computers in Human Behavior},
doi = {10.1016/j.chb.2004.10.006}
}

@article{Oviatt01,
title = {Toward interface design for human language technology: Modality and structure as determinants of linguistic complexity},
journal = {Speech Communication},
volume = {15},
number = {3},
pages = {283-300},
year = {1994},
note = {Special issue on Spoken dialogue},
issn = {0167-6393},
doi = {https://doi.org/10.1016/0167-6393(94)90079-5},
url = {https://www.sciencedirect.com/science/article/pii/0167639394900795},
author = {Sharon L. Oviatt and Philip R. Cohen and Michelle Wang}
}

@article{Litman01,
  title={Spoken versus typed human and computer dialogue tutoring},
  author={Litman, Diane J and Ros{\'e}, Carolyn P and Forbes-Riley, Kate and VanLehn, Kurt and Bhembe, Dumisizwe and Silliman, Scott},
  journal={International Journal of Artificial Intelligence in Education},
  volume={16},
  number={2},
  pages={145--170},
  year={2006},
  publisher={IOS Press}
}

@article{Chen01,
author = {Chen, Qiaochu and Pailoor, Shankara and Barnaby, Celeste and Criswell, Abby and Wang, Chenglong and Durrett, Greg and Dillig, I\c{s}il},
title = {Type-directed synthesis of visualizations from natural language queries},
year = {2022},
issue_date = {October 2022},
publisher = {Association for Computing Machinery},
address = {New York, NY, USA},
volume = {6},
number = {OOPSLA2},
url = {https://doi.org/10.1145/3563307},
doi = {10.1145/3563307},
journal = {Proc. ACM Program. Lang.},
month = oct,
articleno = {144},
numpages = {28}
}

@article{Sah01,
  title={Generating Analytic Specifications for Data Visualization from Natural Language Queries using Large Language Models},
  author={Subham Sah and Rishab Mitra and Arpit Narechania and Alex Endert and John Stasko and Wenwen Dou},
  journal={ArXiv},
  year={2024},
  volume={abs/2408.13391},
  url={https://api.semanticscholar.org/CorpusID:271956821}
}

@article{Maddigan2023Chat2VISFD,
  title={Chat2VIS: Fine-Tuning Data Visualisations using Multilingual Natural Language Text and Pre-Trained Large Language Models},
  author={Paula Maddigan and Teo Susnjak},
  journal={ArXiv},
  year={2023},
  volume={abs/2303.14292},
  url={https://api.semanticscholar.org/CorpusID:257767418}
}

@ARTICLE{Srinivasan04,
  author={Srinivasan, Arjun and Stasko, John},
  journal={IEEE Computer Graphics and Applications}, 
  title={How to Ask What to Say?: Strategies for Evaluating Natural Language Interfaces for Data Visualization}, 
  year={2020},
  volume={40},
  number={4},
  pages={96-103},
  doi={10.1109/MCG.2020.2986902}}

@article{Hoque02,
author = {Hoque, E. and Kavehzadeh, P. and Masry, A.},
title = {Chart Question Answering: State of the Art and Future Directions},
journal = {Computer Graphics Forum},
volume = {41},
number = {3},
pages = {555-572},
doi = {https://doi.org/10.1111/cgf.14573},
url = {https://onlinelibrary.wiley.com/doi/abs/10.1111/cgf.14573},
eprint = {https://onlinelibrary.wiley.com/doi/pdf/10.1111/cgf.14573},
year = {2022}
}

@misc{gu2023systematicsurveypromptengineering,
      title={A Systematic Survey of Prompt Engineering on Vision-Language Foundation Models}, 
      author={Jindong Gu and Zhen Han and Shuo Chen and Ahmad Beirami and Bailan He and Gengyuan Zhang and Ruotong Liao and Yao Qin and Volker Tresp and Philip Torr},
      year={2023},
      eprint={2307.12980},
      archivePrefix={arXiv},
      primaryClass={cs.CV},
      url={https://arxiv.org/abs/2307.12980}, 
}

@inproceedings{Masry2024ChartInstructIT,
  title={ChartInstruct: Instruction Tuning for Chart Comprehension and Reasoning},
  author={Ahmed Masry and Mehrad Shahmohammadi and Md. Rizwan Parvez and Enamul Hoque and Shafiq R. Joty},
  booktitle={Annual Meeting of the Association for Computational Linguistics},
  year={2024},
  url={https://api.semanticscholar.org/CorpusID:268384920}
}

@article{Hay_Duffy_McTeague_Pidgeon_Vuletic_Grealy_2017, title={Towards a shared ontology: A generic classification of cognitive processes in conceptual design}, volume={3}, DOI={10.1017/dsj.2017.6}, journal={Design Science}, author={Hay, Laura and Duffy, Alex H. B. and McTeague, Chris and Pidgeon, Laura M. and Vuletic, Tijana and Grealy, Madeleine}, year={2017}, pages={e7}}

@inproceedings{Prakash01,
author = {Prakash, Yash and Kolgar Nayak, Akshay and Alyaan, Shoaib Mohammed and Khan, Pathan Aseef and Lee, Hae-Na and Ashok, Vikas},
title = {Improving Usability of Data Charts in Multimodal Documents for Low Vision Users},
year = {2024},
isbn = {9798400704628},
publisher = {Association for Computing Machinery},
address = {New York, NY, USA},
url = {https://doi.org/10.1145/3678957.3685714},
doi = {10.1145/3678957.3685714},
booktitle = {Proceedings of the 26th International Conference on Multimodal Interaction},
pages = {498–507},
numpages = {10},
location = {San Jose, Costa Rica},
series = {ICMI '24}
}

@inproceedings{10.1145/3686215.3690154,
author = {Yang, Crystal and Taele, Paul},
title = {An Audiotactile System for Accessible Graphs on a Coordinate Plane},
year = {2024},
isbn = {9798400704635},
publisher = {Association for Computing Machinery},
address = {New York, NY, USA},
url = {https://doi.org/10.1145/3686215.3690154},
doi = {10.1145/3686215.3690154},
booktitle = {Companion Proceedings of the 26th International Conference on Multimodal Interaction},
pages = {46–50},
numpages = {5},
location = {San Jose, Costa Rica},
series = {ICMI Companion '24}
}

@inproceedings{Aziz01,
author = {Aziz, Farkhandah and Creed, Chris and Frutos-Pascual, Maite and Williams, Ian},
title = {Inclusive Voice Interaction Techniques for Creative Object Positioning},
year = {2021},
isbn = {9781450384810},
publisher = {Association for Computing Machinery},
address = {New York, NY, USA},
url = {https://doi.org/10.1145/3462244.3479937},
doi = {10.1145/3462244.3479937},
booktitle = {Proceedings of the 2021 International Conference on Multimodal Interaction},
pages = {461–469},
numpages = {9},
location = {Montr\'{e}al, QC, Canada},
series = {ICMI '21}
}

@inproceedings{Harada01,
author = {Harada, Susumu and Saponas, T. Scott and Landay, James A.},
title = {VoicePen: augmenting pen input with simultaneous non-linguisitic vocalization},
year = {2007},
isbn = {9781595938176},
publisher = {Association for Computing Machinery},
address = {New York, NY, USA},
url = {https://doi.org/10.1145/1322192.1322225},
doi = {10.1145/1322192.1322225},
booktitle = {Proceedings of the 9th International Conference on Multimodal Interfaces},
pages = {178–185},
numpages = {8},
location = {Nagoya, Aichi, Japan},
series = {ICMI '07}
}

@inproceedings{Gorniak01,
author = {Gorniak, Peter and Roy, Deb},
title = {A visually grounded natural language interface for reference to spatial scenes},
year = {2003},
isbn = {1581136218},
publisher = {Association for Computing Machinery},
address = {New York, NY, USA},
url = {https://doi.org/10.1145/958432.958474},
doi = {10.1145/958432.958474},
booktitle = {Proceedings of the 5th International Conference on Multimodal Interfaces},
pages = {219–226},
numpages = {8},
location = {Vancouver, British Columbia, Canada},
series = {ICMI '03}
}

@inproceedings{ROSMI,
author = {Katsakioris, Miltiadis Marios and Konstas, Ioannis and Mignotte, Pierre Yves and Hastie, Helen},
title = {ROSMI: A Multimodal Corpus for Map-based Instruction-Giving},
year = {2020},
isbn = {9781450375818},
publisher = {Association for Computing Machinery},
address = {New York, NY, USA},
url = {https://doi.org/10.1145/3382507.3418861},
doi = {10.1145/3382507.3418861},
booktitle = {Proceedings of the 2020 International Conference on Multimodal Interaction},
pages = {680–684},
numpages = {5},
location = {Virtual Event, Netherlands},
series = {ICMI '20}
}

@article{Luo2025nvBench2A,
  title={nvBench 2.0: A Benchmark for Natural Language to Visualization under Ambiguity},
  author={Tianqi Luo and Chuhan Huang and Leixian Shen and Boyan Li and Shuyu Shen and Wei Zeng and Nan Tang and Yuyu Luo},
  journal={ArXiv},
  year={2025},
  volume={abs/2503.12880},
  url={https://api.semanticscholar.org/CorpusID:277065796}
}

\appendix

\section{Phase I - Exploring Semantic Structure of Chart Creation Instructions}
\begin{figure*}[h]
    \centering
    \includegraphics[width=16cm,keepaspectratio]{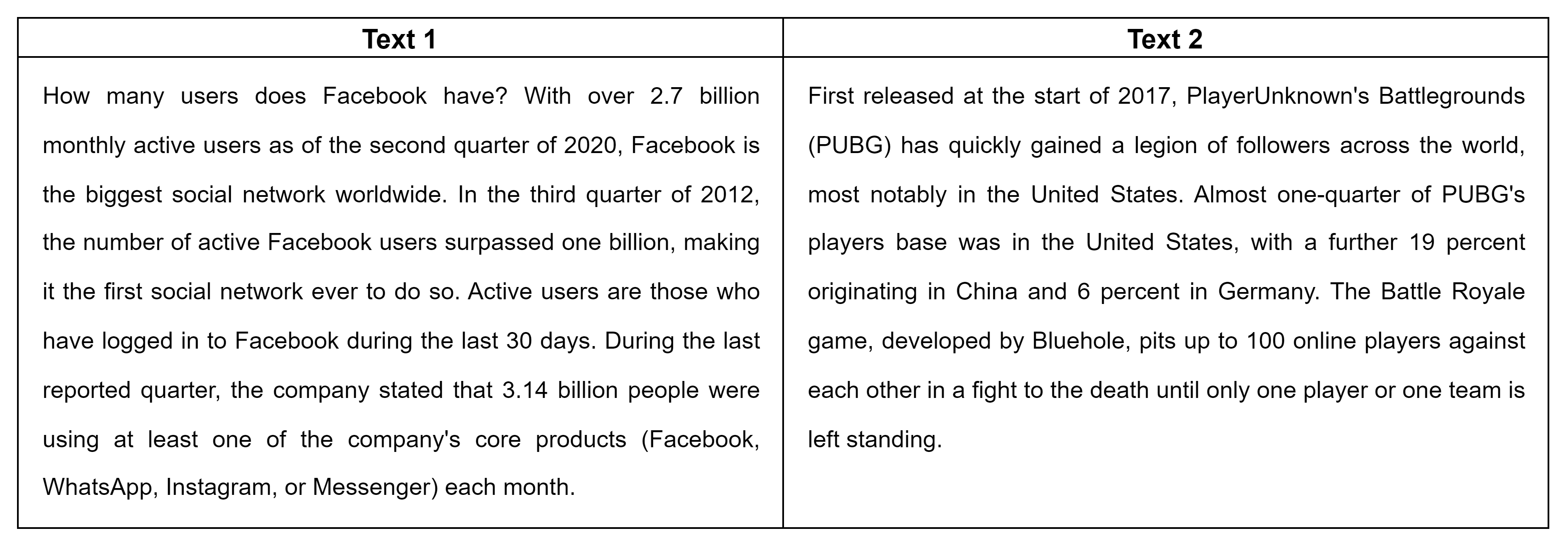}
    \caption{Examples of the text stimuli provided to participants of user studies in Phase I. From \cite{statista}.}
    \label{fig:stimuli_example}
\end{figure*}

\begin{table}[h]
    \centering
    \begin{tabular}{c|c|c|c}
        \textbf{Text} & \textbf{Word Count} & \textbf{Spoken Instructions} & \textbf{Typed Instructions} \\
        \midrule
        Text 1 & 96  & 9  & 13 \\
        Text 2 & 82  & 9  & 9  \\
        Text 3 & 80  & 16 & 9  \\
        Text 4 & 65  & 13 & 11 \\
        Text 5 & 81  & 14 & 15 \\
        Text 6 & 100 & 12 & 16 \\
        Text 7 & 54  & 14 & 11 \\
        Text 8 & 57  & 13 & 8  \\
        \bottomrule
    \end{tabular}
    \caption{Summary of the word count and number of collected spoken and typed imagined‑chart instructions for each text stimulus in Phase I.}
    \label{tab:wordcount-combined}
\end{table}

\begin{figure}[h]
    \centering
    \includegraphics[width=12cm,keepaspectratio]{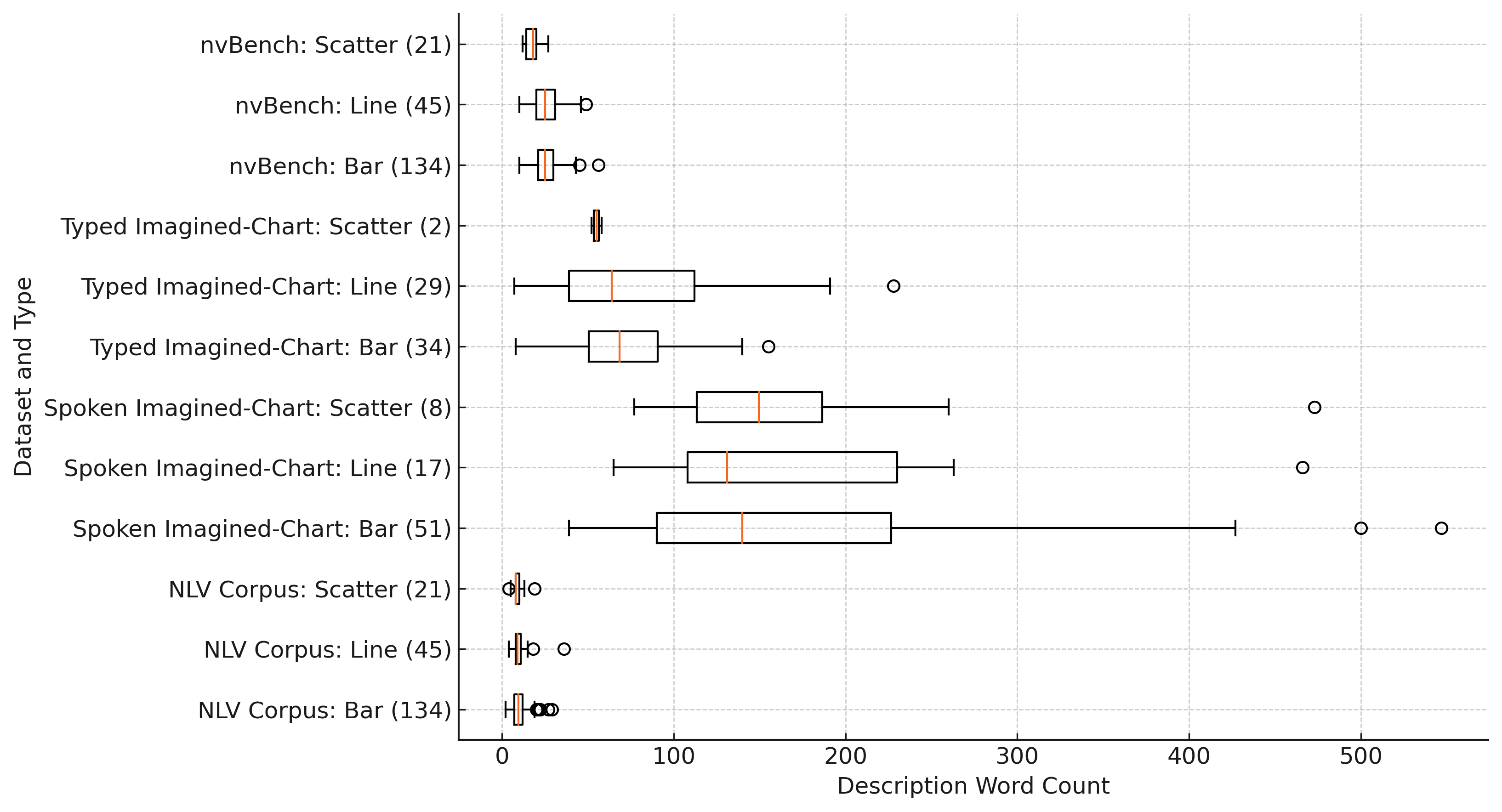}
    \caption{Summary of the word count per instruction of the 76 spoken \newtext{imagined-chart and 65 typed imagined-chart }instructions collected in Phase I, 200 typed \newtext{existing-chart }instructions from the NLV Corpus \cite{Srinivasan01}, and 200 synthetic typed \newtext{existing-chart }instructions from the nvBench \cite{Luo01}. The (count) next to each chart type indicates the total number of instructions associated with that type.}
    \label{fig:word_count_boxplot}
\end{figure}

\section{Phase II - Assessing the Effects of Training Datasets on System Performance}
\begin{figure*}[h]
  \centering
  \includegraphics[width=16cm,keepaspectratio]{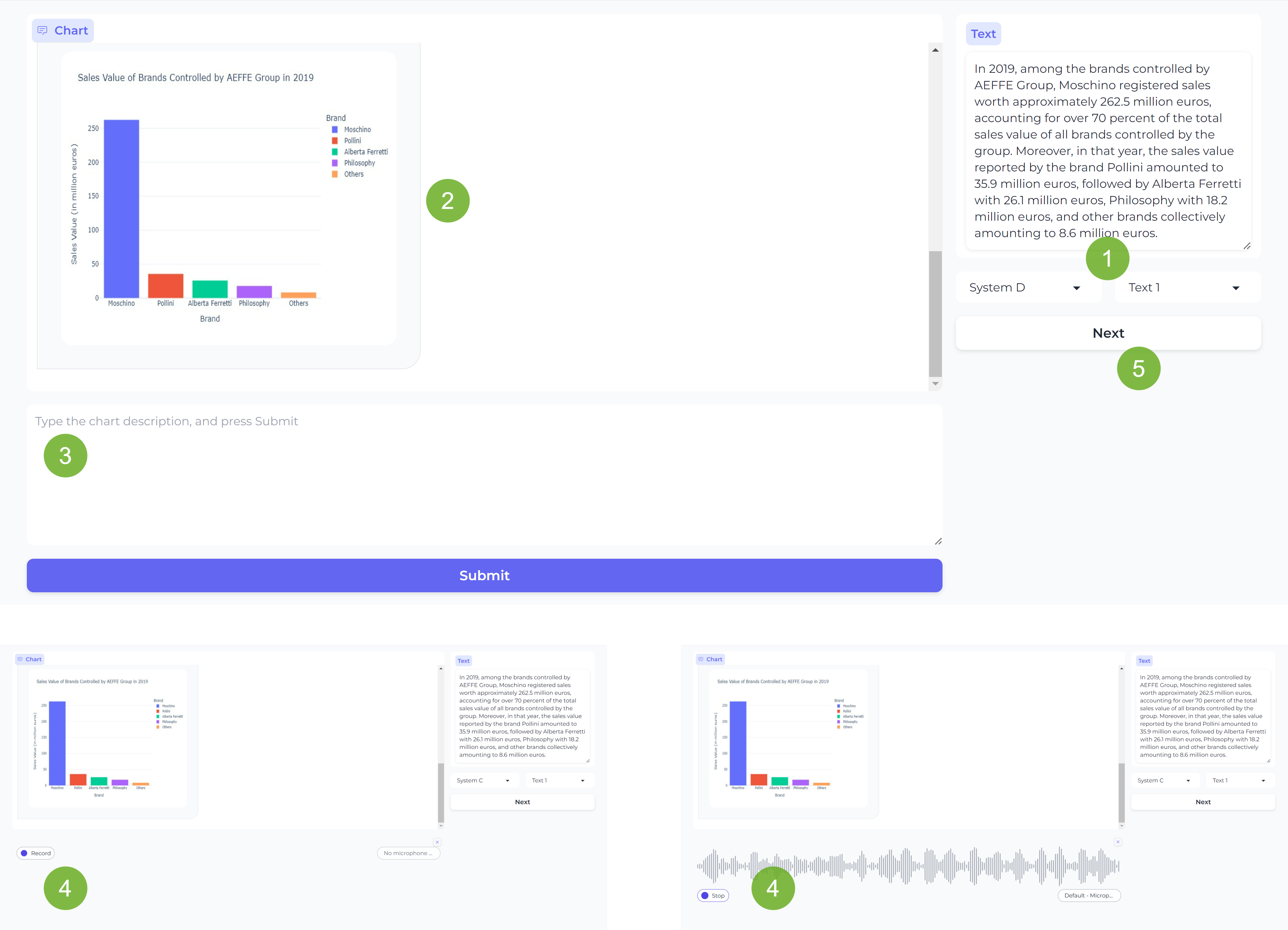}
  \caption{User's view of the web-based study environment in Phase II. The interface layout has five sections: the information panel with the current system name and text stimulus (1), the chart display area with the history of provided instructions and generated charts (2), the input field for typing (3), the recording start and stop button (4), and the navigation button to clear the history and proceed to the next text stimulus (5).}
  \label{fig:interface02}
\end{figure*}

\textit{Interface workflow:} For text input, the user can type text instructions in the text field and press the submit button. For spoken input, the user can click a record button to start and end dictating the voice description of the chart using a microphone. 
The recording is then automatically transcribed into text using an automatic speech recognition system. While the current prototype only supports descriptions in English, the possibility of extrapolating it to other languages exists with the emergence of robust multilingual voice recognition models. 
Transcribed audio and text input is sent for processing to the corresponding system, and the generated visualization code is executed and displayed in the chart display area. 
Users can make additional edits by recording or typing a new instruction. The instruction, together with the previous history of instructions and generated charts, will be again sent to the corresponding system, and the generated visualization code is again executed and displayed. Once the user finishes, they can click the navigation button to move to the next text stimulus.

\begin{figure*}[h]
    \centering
    \includegraphics[width=16cm,keepaspectratio]{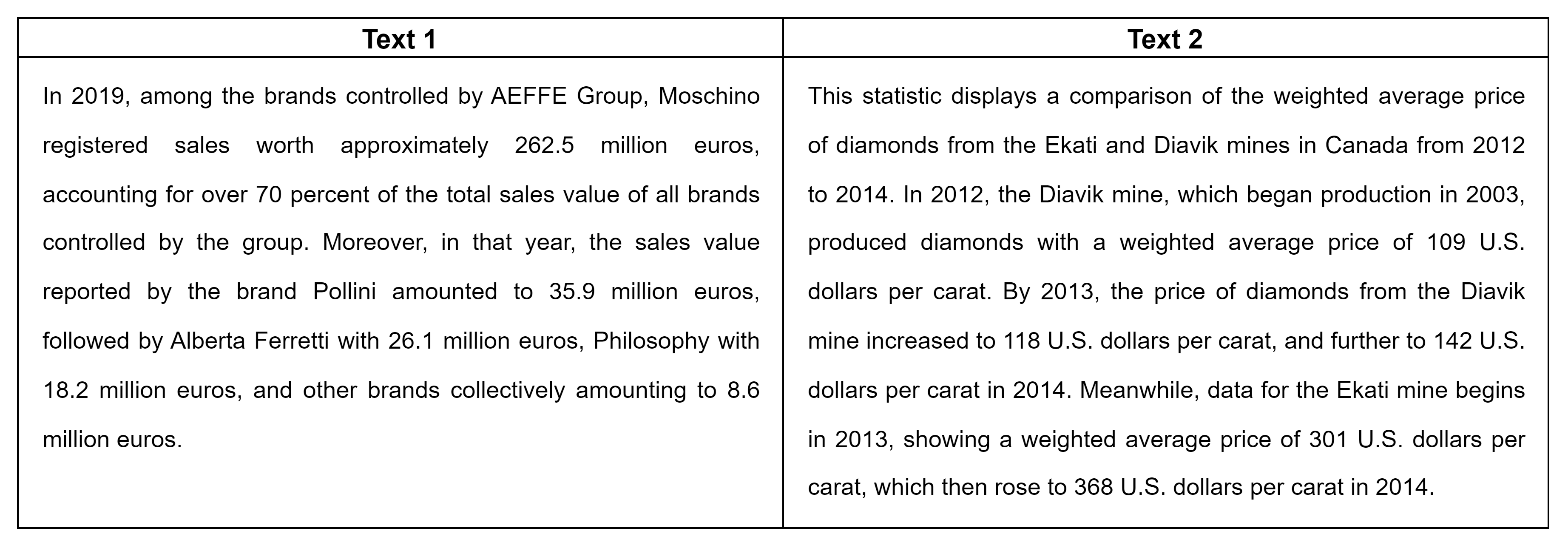}
    \caption{Examples of the text stimuli provided to participants of the evaluation study in Phase II. From \cite{statista}.}
    \label{fig:stimuli_example_2}
\end{figure*}

\begin{table}[h]
    \centering
    \begin{tabular}{c|c|c}
        \textbf{Text} & \textbf{Word Count} & \textbf{Collected Instructions} \\
        \midrule
        Text 1 & 73 & \multirow{8}{*}{19} \\
        Text 2 & 105 &  \\
        Text 3 & 74 &  \\
        Text 4 & 76 &  \\
        Text 5 & 69 &  \\
        Text 6 & 69 &  \\
        Text 7 & 66 &  \\
        Text 8 & 103 &  \\
        \bottomrule
    \end{tabular}
    \caption{Summary of the word count and number of collected instructions for each text stimulus in Phase II.}
    \label{tab:wordcount_2}
\end{table}

\end{document}